\documentclass[12pt,preprint]{aastex}

\slugcomment{Accepted to ApJ (November, 28 2012)}

\shorttitle{Hierarchical Fragmentation of the Orion Molecular Filaments.}
\shortauthors{Takahashi et al.}

\begin{document}

\title{Hierarchical Fragmentation of the Orion Molecular Filaments}

\author{Satoko Takahashi$^{1}$, Paul T. P. Ho$^{1, 2}$, Paula S. Teixeira$^{3,4}$, Luis A. Zapata$^{5}$, Yu-Nung Su$^{1}$}
\affil{$^{1}$Academia Sinica Institute of Astronomy and Astrophysics, P.O. Box 23-141, Taipei 10617, Taiwan; $satoko{\_}t@asiaa.sinica.edu.tw$, \\ 
$^{2}$Harvard-Smithsonian Center for Astrophysics, 60 Garden Street Cambridge, MA 02138, U.S. A., \\
$^{3}$Universitaet Wien, Institut fuer Astrophysik, Tuerkenschanzstrasse 17, 1180, Wien, Austria, \\
$^{4}$Laborat\'orio Associado Instituto D. Luiz-SIM, Universidade de Lisboa, Campo Grande, 1749-016, Lisbon, Portugal \\
$^{5}$Centro de Radioastronom\'\i a y Astrof\'\i sica, Universidad Nacional Aut\'onoma de M\'exico, Morelia, Michoac\'an 58090, M\'exico}

\begin{abstract} 
We present a high angular resolution map of 850 $\mu$m continuum emission of the Orion Molecular Cloud-3 (OMC 3) 
obtained with the Submillimeter Array\footnote{The Submillimeter Array is a joint project between the Smithsonian 
Astrophysical Observatory and the Academia Sinica Institute of Astronomy and Astrophysics and is funded 
by the Smithsonian Institution and the Academia Snica.} (SMA); 
the map is a mosaic of 85 pointings covering an approximate area of \hbox{$6'.5{\times}2'.0$} (0.88$\times$0.27 pc). 
We detect 12 spatially resolved continuum sources, each with an \hbox{H$_2$} mass between \hbox{0.3--5.7 M$_{\odot}$} 
and a projected source size between \hbox{1400--8200 AU}.  
All the detected sources are on the filamentary main ridge ($n_{\rm{H_2}}{\geq}10^6$ cm$^{-3}$), and analysis based 
on the Jeans theorem suggests that they are most likely gravitationally unstable. 
Comparison of multi-wavelength data sets indicates that of the continuum sources, 
6/12 (50 \%) are associated with molecular outflows, 8/12 (67 \%) are associated with infrared sources, 
and 3/12 (25 \%) are associated with ionized jets. The evolutionary status of these sources ranges 
from prestellar cores to protostar phase, confirming that \hbox{OMC-3} is an active region with ongoing 
embedded star-formation. 
We detect quasi-periodical separations between the OMC-3 sources of $\approx$17$''$/0.035 pc. 
This spatial distribution is part of a large hierarchical structure, that also includes fragmentation scales of GMC 
($\approx$35 pc), large-scale clumps ($\approx$1.3 pc), and small-scale clumps ($\approx$0.3 pc), 
suggesting that hierarchical fragmentation operates within the Orion A molecular cloud. 
The fragmentation spacings are roughly consistent 
with the thermal fragmentation length in large-scale clumps, while for small-scale cores it is smaller than 
the local fragmentation length. These smaller spacings observed with the SMA can be explained by either a helical 
magnetic field, cloud rotation, or/and global filament collapse. Finally, possible evidence for sequential 
fragmentation is suggested in the northern part of the OMC-3 filament. 
\end{abstract}
\keywords{ISM: clouds --- ISM: individual (OMC-3) ---stars: formation --- Stars: protostars ---
ISM: structure --- ISM: individual (Orion A)}

\section{INTRODUCTION}
The key steps of the star formation process such as large-scale gas infall, disk accretion, 
and mass ejection have been studied mainly in nearby isolated star-forming regions by various 
methods such as the analysis of: Spectral Energy Distributions (SEDs), gas kinematics, density 
structures of parental cores, and molecular outflows/jets (e.g., Evans et al. 1999; Bachiller \& Tafalla 1999; 
Myers et. al. 2000 and therein). 
However, infrared studies suggest that the majority of stars \hbox{(80--90\%)} form as members 
of embedded clusters (N${\ast}{\geq}$35) within nearby giant molecular clouds (GMCs; located at $d{\leq}2$ kpc), 
rather than in isolated star formation environments (e.g., Lada \& Lada 2003; Porras et al. 2003). 
Hence, studies of the formation and evolutionary processes of embedded clusters, as well as associated GMCs, 
are crucial to understand the general features of star formation. 
Moreover, these embedded clusters, identified as a group of infrared excess sources, are often associated 
with dense filamentary structures \hbox{(i.e., N$_{H_2}>$10$^{22}$ cm$^{-2}$)}, and most likely represent 
the youngest association of the embedded stellar-clusters (Myers 2009). 
In order to observationally investigate the initial conditions of embedded clusters, 
it is essential to spatially resolve the individual star-forming cores within molecular filaments, 
obtain kinematic information, and investigate the fragmentation properties and physical conditions of each of the cluster members. 

The Orion A molecular cloud (\hbox{$d=414$ pc}; Sandstrom et al. 2007; Menten et al. 2007; Hirota et al. 2007; Kim et al. 2008), 
which contains a large amount of gas \hbox{(${\sim}5{\times}10^4~M_{\odot}$)}, is the nearest GMC, and one of 
the best-studied embedded cluster-forming regions at all observable wavelengths (e.g., Bally et al. 1987; 
Allen et al 2007; Tatematsu et al. 1993, Johnstone \& Bally. 1999; Tsujimoto et al. 2005; O'Dell et al. 2008). 
A large number of molecular line observations have been performed: in \hbox{$^{12}$CO} by Maddalena et al. (1986), 
Castets et al. (1990), in \hbox{$^{13}$CO} by Bally et al. (1987); Buckle (2012), 
in \hbox{C$^{18}$O }by Dutrey et al. (1991; 1993); Buckle (2012), in CS by Tatematsu et al. (1993), in NH$_3$ 
by Castets et al. (1990), Harju et al. (1991), Cesaroni \& Wilson (1994), and Wiseman \& Ho (1998), 
and in \hbox{H$^{13}$CO$^{+}$} by Aso et al. (2000), Ikeda et al. (2007), Tatematsu et al. (2010). 
These observations revealed the distribution of the dense gas within the Orion A GMC down to 0.1 pc scale. 
In addition to molecular line observations, (sub)millimeter continuum observations performed by single-dish 
telescopes have revealed a remarkable filamentary structure over the entire Orion A molecular cloud with 
a length of \hbox{$\sim$7 pc}. A large number of dust continuum sources, candidate \hbox{Class 0/I} sources, 
were detected, and properties of the dust emission along the Orion Molecular Cloud (OMC) filaments were 
investigated (Chini et al. 1997; Lis et al. 1998; Johnstone \& Bally 1999; Nielbock et al. 2003; Nutter \& Ward-Thompson 2007). 
Furthermore, near- to mid- infrared surveys by {\it Spitzer} \hbox{MIPS/IRAC} showed that a number of infrared excess sources (YSO candidates) 
are actually located on the OMC filaments with a wide stellar mass range, from brown dwarfs to massive stars (Nielbock et al. 2003; Allen et al. 2007;  
Peterson et al. 2008; Peterson \& Megeath 2008; Sadavoy et al. 2010; Megeath et al. 2012). 
Moreover, recent high-angular resolution radio observations, targeted at relatively bright (sub)millimeter sources 
(i.e., intermediate-mass protostellar candidates), directly resolved detailed spatial structure of protostellar cores 
and associated molecular outflows in the OMC filaments. The detected sources showed a variety of properties such as different 
structures, kinematics, and evolutionary phases (e.g., Williams et al. 2003; Zapata et al. 2004; 2005; 2006; 2007; Matthews et al. 2005; 
Takahashi 2006; 2008a; 2008b; 2009; 2012; Shimajiri et al. 2008; 2009; Takahashi \& Ho 2012). 
However, so far only a very limited number of millimeter and submillimeter observational studies have been made 
which achieve an angular resolution comparable to, or better than, existing IR data (i.e., \hbox{$\leq$a few arcsec resolution}, 
which can spatially resolve \hbox{$\leq$1000 AU} scale structures at the distance of Orion). Hence, the physical relation between 
IR sources and the small-scale gas structure ($D{\leq}$ a few 1000 AU and high-density gas: n$_{H2}{\geq}10^5$ cm$^{-3}$), 
which can be spatially resolved with the millimeter/submillimeter interferometric observations, remains unclear. 
It is also important to spatially resolve individual star forming sites, or cores, in order to study the (IMF precursor) core mass function (CMF). 
Analysis of the CMF could lead to the understanding of the similarities and differences in the IMF of entire 
stellar-clusters which may have been imposed at the initial evolutionary stage. 

In order to observationally study the nature of embedded sources and their associated filaments, as a first step, we report 
\hbox{850 $\mu$m} continuum observations of the molecular filaments toward the OMC-3 region located in the northern part 
of Orion A GMC with the largest mosaic observations ever published using the SMA \hbox{($6'.5{\times}2'.0$; 0.88$\times$0.27 pc)}. 
The SMA, equipped with \hbox{345 GHz} receivers is one of the most powerful facilities to spatially resolve prestellar/protostellar 
cores and circumstellar disks through the thermal dust continuum emission. The goals of the paper are to (i) spatially resolve 
the individual star-forming cores  with the angular resolution of 4$''$.5 ($\approx$1900 AU) and identify the prestellar and 
protostellar candidates with a mass sensitivity of \hbox{0.15 M$_{\odot}$}, (ii) compare the physical properties of 
(sub)millimeter sources with existing multi-wavelength data and study the evolutionary status of each identified source, 
and (iii) quantify the physical properties of the filaments such as fragmentation scale lengths by comparing our SMA images 
with previous single-dish images.

We explain in Section 2 the strategies of the observations, data reduction, and imaging. 
We present in Section 3 the synthesized images and the derived physical parameters of the filaments. 
In Section 4, we discuss the physical properties and the evolutionary status of identified cores and 
the derived fragmentation properties of the OMC filaments. 
Finally, we summarize the paper and discuss future prospects in Section 5. 

This paper serves as a starting point for a series of high-angular resolution OMC studies focused on the filamentary 
structures and their embedded sources. Our studies, which will cover the entire system of OMC filaments in the continuum, 
as well as molecular line emissions in the (sub)millimeter (described in Section 2.1), will include a statistically significant number 
of embedded sources. Once the Atacama Large Millimeter/submillimeter Array (ALMA) and the Jansky Very Large Array (JVLA) are fully operational, 
the sensitivity and resolution of their data will be greatly improved so that the wide-field mosaic 
observations will be a common approach, especially for GMCs and nearby extragalactic objects. 
Hence our studies would be very important in developing a strategy to investigate structures of molecular clouds and associated proto-clusters 
on pc-scale down to AU scale.

\section{OBSERVATIONS AND DATA REDUCTION}
	\subsection{SMA Observations}
	
		The observations were carried out at 351 GHz using the eight \hbox{6 m} antennas of the SMA (Ho et al. 2004) 
		in the compact configuration on January 2 and October 21, 2008 and in the subcompact configuration on August 29, 2008. 
		The primary beam at this frequency is \hbox{36$''$}, and the 85 pointing centers were arranged with sampling in a nyquist 
                grid spaced by \hbox{18$''$} with 
		a position angle of \hbox{45$^{\circ}$} (Figure 1). We observed each field for one minute during each of \hbox{4 -- 18} 
                cycles through all the fields, 
		giving a total on-source integration time of \hbox{4--18 minutes}. 
		Both the LSB \hbox{(${\nu}_c$=346 GHz)} and USB \hbox{(${\nu}_c$=356 GHz)} data were obtained 
		simultaneously with the \hbox{90$^{\circ}$} phase switching technique by the digital spectral correlator, 
		which had a bandwidth of \hbox{2 GHz} for each sideband. 
		
		The SMA correlator covers \hbox{2 GHz} bandwidth in each of the two side-bands separated by \hbox{10 GHz}. 
		Each band is divided into \hbox{24 chunks} of \hbox{104 MHz} width, which were covered by fine spectral 
		resolution (\hbox{256 channels} correspond to a velocity resolution of \hbox{0.36 km s$^{-1}$} in our observing setting). 
		In addition to continuum observations, molecular lines such as \hbox{HCO$^{+}$(4--3)}, \hbox{H$^{13}$CO$^{+}$(4--3),} 
		\hbox{CO(3--2),} \hbox{SiO(8--7),} and several \hbox{CH$_3$OH} and SO transitions were simultaneously obtained 
		with the \hbox{2 GHz} band width in each sideband. These molecular line data will be presented in a future article.
		
		The continuum data from both sidebands were combined to obtain a higher \hbox{signal-to-noise} ratio. 
		The combined configurations of the arrays provided projected baselines ranging from \hbox{9.7 to 88 k${\lambda}$}. 
		Our observations were insensitive to structures larger than 17$''$ (7000 AU) at the 10\% level (Wilner \& Welch 1994). 
		Passband calibration was achieved by observing the quasars, \hbox{3C 273} and \hbox{3C 279}. 
		Amplitudes and phases were calibrated by observations of \hbox{0530+135} and \hbox{0423-013}, 
		whose flux densities were determined relative to Titan and Neptune. The overall flux uncertainty was estimated to be \hbox{$\sim$20\%}. 
		The pointing accuracy of the SMA observations was \hbox{${\sim}3''$}. 
		The raw data were calibrated using MIR, originally developed for the Owens Valley Radio Observatory (Scoville et al. 1993) 
		and adopted for the SMA. 
		The SMA observational log for each date and the observational parameters are summarized in Table 1 and Table 2. 
		
	\subsection{Mosaic Imaging Procedures}	
		In order to produce the mosaic continuum image, certain frequencies (which include possible molecular lines described above) 
		are subtracted from the visibility data. 
		The effective bandwidth for the continuum emission is approximately 3.2 GHz. 
		CLEANed images in each pointing were made using 
		DIFMAP with a Gaussian taper \hbox{(HWHM=25 k${\lambda}$)}.  
		Due to residual phase errors, it was difficult to remove the increased noise, which is greater than 5$\sigma$ level, around 
		the four brightest compact sources, \hbox{SMM 2,} \hbox{SMM 3,} \hbox{SMM 8,} and \hbox{SMM 11}. 
		These enhanced noise structures are due to 
		the limited dynamic range which can be achieved against the strong peak signal levels of those bright sources. 
		In order to improve the dynamic range of the images, we reduced the phase noise errors, by applying a self-calibration procedure toward 
		the several fields which include either \hbox{SMM 2,} \hbox{SMM 3,} \hbox{SMM 8,} or \hbox{SMM 11} 
		within their primary beams. 
		Within the integration time of \hbox{4--18 min}., the theoretical noise levels were estimated to be \hbox{13-- 25 mJy beam$^{-1}$}. 
		After self-calibration, the theoretical noise levels were achieved around \hbox{SMM 2,} \hbox{SMM 3,} and \hbox{SMM 11}. 
		Images around \hbox{SMM 8} (i.e., nine mosaic points centered on \hbox{SMM 8}) achieved \hbox{1.1 to 2.0} times theoretical noise levels. 
		This corresponds to a dynamic range of \hbox{$\sim$200} around \hbox{SMM 8}.  
				
		After the CLEANing and self-calibration processes, the 85 images were imported into Miriad (Wright \& Sault 1993) for primary beam corrections, 
		and then combined into a single image using the miriad task \hbox{``linmos''}. 
		Since each image had slightly different angular resolution, 
		all the images were convolved to the final beam size of \hbox{4.5$''$} FWHM (corresponding to \hbox{1900 AU}). 
		The achieved rms noise level for the entire mosaic image was estimated to be \hbox{20 mJy} on average.

\section{RESULTS} 
	\subsection{Identification of SMA \hbox{850 $\mu$m} Compact Continuum Sources}
		
		Figure 2 shows the SMA 850 $\mu$m continuum in the left panel, and in the right panel we compare the SMA image with 
		the previously published JCMT/SCUBE 850 $\mu$m image by Johnstone \& Bally (1999). 
		We clearly detected 12 compact SMA continuum sources embedded within the filament with 
		a \hbox{signal-to-noise} ratio of greater than 5 \hbox{(flux more than 100 mJy beam$^{-1}$}; 
		denoted by white contours in the right panel of \hbox{Figure 2}). 
		All the detected continuum sources are found along the main ridge of the OMC-3 filament, with no detections off the main ridge. 
		This indicates that the relatively massive sources in this region are likely formed only within the densest part of the filaments.  
		In order to measure the position, peak flux, total flux density, and deconvolved size of the continuum source, 
		2D Gaussian fitting in AIPS (task $``$JMFIT$''$) was used. The measured results and their uncertainties 
		are listed in Table 3. The spatial structures of the individual sources are presented in \hbox{Figure 3}. 
		
		Eight out of the twelve SMA continuum peaks (\hbox{SMM 1--6,} \hbox{SMM 7,} and \hbox{SMM 10}) coincide with 
		(sub)millimeter continuum sources previously detected with the JCMT/SCUBA, CSO/SHARC, and 
		IRAM/MPIfR, to within the single-dish positional accuracies of ${\sim}{\pm}5''$ 
		(Chini et al. 1997, Lis et al. 1998, Johnstone \& Bally 1999). 
		\hbox{SMM 10} was detected in the previous 3.3 mm interferometric observations at the \hbox{3$\sigma$} 
                level as \hbox{MMS 7-NE} 
		(\hbox{Figure 2a} of Takahashi et al. 2006). 
		Figure 1 of Takahashi et al. (2009) shows an image where SMM 7 (MMS 6-NE) is spatially resolved using 
                SMA compact configuration data with robust weighting. 
		The sources, \hbox{SMM 9} and \hbox{SMM 12,} were  detected for the first time with our SMA 850 $\mu$m continuum observations. 
		For the other eight sources, this is the first time that they have been spatially resolved and positioned accurately. 
		On the assumption that the \hbox{850 $\mu$m} emission comes from optically thin dust emission 
		and that the temperature distribution of the dust continuum emission is uniform, we can estimate the source mass by 
		
		\begin{equation}
			M_{\rm{H_2}}={\frac{F_{\lambda}d^2}{{\kappa}_{\lambda}B_{\lambda}(T_{\rm{dust}})}},
		\end{equation}
		
		\noindent
		where $\kappa_{\lambda}$ is the mass-absorption coefficient of dust grains, $B_{\lambda}(T)$ is 
                the Planck function for a dust temperature $T_{\rm{dust}}$, 
		$F_{\lambda}$ is the total flux density of the continuum emission, and $d$ is the distance to the source (414 pc; Menten et al. 2007). 
		Adopting a dust opacity of \hbox{$\kappa_{\lambda}$=0.037 cm$^{2}$ g$^{-1}$ (${\lambda}$/400 $\mu$m)
		$^{-{\beta}}$} (Keene et al. 1982), \hbox{${\beta}=1.6$} (Johnstone \& Bally 1999), a dust temperature of \hbox{$T_{\rm{dust}}=20$ K} 
		(Cesaroni \& Wilson 1994, Chini et al. 1997; Tatematsu et al. 2008), \hbox{a gas-to-dust ratio=100}, 
                and measured total flux densities (Table 3), 
		the masses of the continuum sources are estimated to range from  \hbox{0.3 to 5.7 $M_{\odot}$}. 
		High mean molecular hydrogen volume densities were also derived toward these sources with a range between  
		\hbox{3.2$\times$10$^6<n_{\rm{H_2}}<$3.3$\times$10$^8$ cm$^{-3}$} under the assumption of a spherical
		geometry \hbox{($n_{\rm{H_2}}=M_{\rm{H_2}}/({\frac{4}{3}}{\pi}{r^3}{\mu}{m_{\rm{H}}})$)}. 
		Here, $\mu$ and $m_H$ are the hydrogen mean molecular weight (2.33) and the hydrogen mass, respectively, 
		and $r$ is the radius of the source, taken as the geometric mean of the major and minor axes of its deconvolved size. 
		The derived physical parameters of these sources are summarized in Table 4. 
				
		Note that the total flux of the 850 $\mu$m continuum emission might include contamination from ionized emission, 
                which comes from free-free continuum jets of protostars. 
		From \hbox{3.6 cm} observations (Reipurth et al. 1999), and  with the assumption 
		that the typical spectral index of free-free jets is 0.6 (Anglada et al. 1998; Reynolds 1986), we find that 		 
		this contamination at \hbox{850 $\mu$m} is negligible (i.e., \hbox{$\leq$0.4 \%}) for all the detected OMC-3 sources. 
		The continuum emission at \hbox{850 $\mu$m} must therefore come from the thermal dust.

		In order to estimate the flux concentration factors, \hbox{$C_{\rm{(SMA/JCMT)}}$}, the SMA data were first convolved to the JCMT 14$''$ beam, 
		and then the peak intensity within the JCMT 14$''$ beam (Jy beam$^{-1}$) are used for deriving the ratios. 
		The flux concentration factor for \hbox{SMM 8} (or \hbox{MMS 6-main;} Takahashi et al. 2009;2012b) is 86\%, 
		approximately twice higher than those of the other detected continuum sources 
		(The mean and median value of the flux ratios were estimated to be \hbox{41\%} and \hbox{35 \%}, respectively). 
		\hbox{SMM 8} shows small deconvolved sizes compared with most of the other SMA sources and has the highest flux 
                among the 12 detected sources (\hbox{Table 4}).  
		No clear difference of the flux concentration ratio was detected between the prestellar cores 
		(i.e., core not associated with any outflows nor any IR sources; Table 5) and the protostellar cores 
		(i.e., core associated with either outflows or IR sources; Table 5) except for the higher ratio value derived in SMM 8. 
		The evolutionary stage of detected sources will be discussed in \hbox{Section 4.2}. 
		
		Moreover, to determine the flux-recovering ratio for the entire filament, we proceeded in the following manner. 
		First, we calculated the total SMA flux over the entire image (after convolving the image with the JCMT beam), 
                $F_{\rm{SMA}}$ ($\approx$16 Jy/14$''$ beam); 
		this corresponds to the sum of the fluxes of each of the SMA sources since we detected no extended emission 
		from the filament itself in our SMA image. Second, we determined the JCMT contour level that corresponds to 
		the the SMA 5$\sigma$ detection level (100mJy/4$''$.5 beam), i.e., 0.97Jy/14$''$ beam; 
		this was used as a lower threshold for extended emission in the JCMT image. 
		We then summed up the flux within this contour in the JCMT image, $F_{\rm{JCMT}}$ ($\approx$103 Jy/14$''$ beam). 
		Finally, we divided the two fluxes, $F_{\rm{SMA}}/F_{\rm{JCMT}}$, and found that only 16\% of the total JCMT flux 
                was recovered in our SMA data.		
			
		It is technically possible to combine the SMA \hbox{850 $\mu$m} data with the \hbox{JCMT/SCUBA} data 
		in order to recover the flux from short antenna spacings.  
		However, we did not apply this method due to the following reasons: (i) JCMT/SCUBA is a bolometer and covers a much 
		wider frequency range as compared to the SMA (${\Delta}B_{\rm{JCMT}}{\gtrsim}7{{\Delta}B_{\rm{SMA}}}$),  
		and (ii) JCMT/SCUBA frequency coverage includes several strong molecular lines such as CO, HCO$^{+}$, and SO. 
		Hence, the JCMT continuum emission map may be contaminated by emission from these molecular lines.

	\subsection{Physical Properties of the Continuum Sources}
		In order to further discuss the properties of the detected sources, it is important to know the limits of our SMA observations. 
				
		Figure 4(A) shows the mass distribution of the detected sources in OMC-3 (filled histogram). 
		The median and mean value of the source masses were estimated to be 1.6 M$_{\odot}$ and 1.9 M$_{\odot}$, respectively. 
		Among the detected sources, \hbox{SMM 8} (or \hbox{MMS 6-main} in Takahashi et al. 2009) is a source which has a gas mass 
		3.5 times greater than the median value.
		Note that \hbox{SMM 8} has the largest mass, the largest flux concentration factor, and one of the most compact source sizes, 
		indicating a very different nature as compared to the other sources in OMC-3. Detailed properties of this source are discussed 
		in Takahashi et al. (2009); (2012), and Takahashi \& Ho (2012).  
	
		Since the number of detected sources were limited, and the error bars for the distribution scale as $\sqrt{N}$ 
		(a Poisson distribution was adopted for the errors), 
		it is difficult to discuss the characteristic mass from our present results. 
		A full analysis of core census will be our future goal with a more extensive sample.
		However, an interesting question is whether we can extrapolate the mass distribution below the detection limit of our observations. 
		According to our previous calculation, the JCMT ``residual flux'' is $F_{\rm{JCMT}}-F_{\rm{SMA}}$=87 Jy. 
		This total flux corresponds to an H$_2$ mass of 98 M$_{\odot}$, with an assumption of \hbox{$\beta$=1.6} (Johnstone \& Bally 1999).  
		This suggests that the residual gas has the potential to form numerous cores (\hbox{$\sim$98} of 0.1 M$_{\odot}$ cores) 
		even with a \hbox{10 \%} star formation efficiency (estimated from the nearby embedded clusters; e.g., Lada \& Lada 2003). 
		
		We consider that it is possible to detect sources less than \hbox{0.1 M$_{\odot}$} 
		because recent SMA \hbox{1.3 mm} continuum observations have actually detected 31 circumstellar disks 
		with a mass range from \hbox{0.0028 -- 0.0662 M$_{\odot}$} toward the same Orion star-forming region 
		(28 in the Trapezium cluster and 3 outside of the Trapezium cluster; Mann \& Williams 2009 a, b) as shown in \hbox{Figure 4A} 
                (open histogram). 
		The relatively shallow and large mosaic observations we performed, as compared to the targeted observations performed by 
                Mann \& Williams (2009 a, b), make 
		it difficult to detect these less massive circumstellar matter around the relatively evolved young stars (Class II sources).  
		Moreover, note that these disks have much smaller sizes ($<$300 AU) than that of the sources detected in our SMA observations. 
		
		Figure 4(B) shows the source size distribution. The source sizes are mainly distributed from 1000 to \hbox{4000 AU}.
		In our OMC-3 sample, the minimum detectable source size is roughly limited by the present SMA angular resolution
		\footnote{The minimum detectable size is also related to the signal-to-noise ratio of the detected sources. Since enough 
                 signal-to-noise ratio ($\geq$50) has 
		been achieved in our observations, we have succeeded to spatially resolve the structures less than the synthesized beam.}.
		However, as noted before, the results of the Orion disk survey suggest the presence of more compact sources 
		(\hbox{$D<$ 300 AU}; Vicente \& Alves 2005; Mann \& Williams 2009 a, b). 
		The maximum detectable source size is determined by the shortest baseline in the database, which is 
                \hbox{${\lambda}/{D_{\rm{min.}}}{\sim}19''$} or \hbox{7800 AU}. 
		Further discussion on the detection of extended sources can be found in Wilner \& Welch (1994). 
		They used Gaussian models with limited $uv$ sampling. 
		If we assume Gaussian shaped sources, our observations are sensitive to \hbox{$\sim$17$''$} (7000 AU) structures at the 
                \hbox{10 \%} level.
		The source size of SMM 5, shown as the largest circular source size in Figure 4(B), is larger than the SMA detectable size --  
		this is because we used a two-dimensional Gaussian fitting method for determining the source size. SMM 5's weaker 
                peak intensity (compared to other sources), 
		and its non-Gaussian structure, resulted in a two-dimensional Gaussian fit with a large FWHM. 
		
		In the residual image, the mean flux is \hbox{1.4 Jy} within the \hbox{JCMT 14$''$} beam. 
		With the assumption of uniform flux distribution, the mean flux within the SMA \hbox{4$''$.5} beam is estimated to be \hbox{144 mJy}. 
		This value is larger than the SMA \hbox{5$\sigma$} level (100 mJy), which is the threshold for source identification.
		However, unless this missing flux arises from residuals that are smaller than the maximum detectable size, it will not be detected. 
		Therefore, the missing flux most likely arises from the more extended filamentary structure (${\geq}19''$).

\section{DISCUSSION}		

	\subsection{Star formation in OMC-3}	
		As shown in the previous section, potential star forming cores have been spatially resolved at \hbox{850 $\mu$m} with the SMA.
		The origin of the \hbox{850 $\mu$m} continuum emission comes mostly from thermal dust emission, suggesting dense gas associated with circumstellar envelopes.
		Physical properties of the detected continuum sources and the star formation activities in the OMC-3 region are discussed in the following subsections. 
				
		\subsubsection{Gravitational Stability of the Continuum Sources}
		
		In order to discuss the physical properties of the detected SMA continuum sources, their measured source sizes are plotted as a function of the estimated 
		hydrogen number density. Jeans analysis can provide a useful way to discuss the gravitational stability of the observed sources. 
		With the assumption of an infinite and homogenous medium, the Jeans length is described as follows (Jeans 1902),  
		
		\begin{equation}
			\lambda_{\rm{frag.}} = \sqrt{ \frac{{\pi}c_s^2}{G{\rho_0}}}
		\end{equation}
		
		\noindent
		where G is the gravitational constant, $\rho_0$ is the mean density, and $c_s$ is the sound speed  
		(the relation between the sound speed and the temperature of the gas are described as $c_s = \sqrt{\frac{kT_{\rm{gas}}}{{\mu}m_H}}$), 
		where $T_{\rm{gas}}$ is the gas temperature and $k$ is the Boltzman constant. 
		The Jeans length corresponds to the critical scale when the gas cloud becomes gravitationally unstable, 
		i.e., if the source size is larger than the Jeans length ($D_{\rm{core}}{>}{\lambda}_{\rm{frag.}}$), the core becomes gravitationally unstable 
		and starts to collapse. 
		The Jeans critical number density is described as a function of the Jeans length and gas temperature as:
				
		\begin{equation}
			n_{\rm{H_2}} (\lambda_{\rm{frag.}}, T_{\rm{gas}}) = \frac{{\pi}k}{G({\mu}{m_H})^2} \frac{T_{\rm{gas}}}{\lambda_{\rm{frag.}}^2}
		\end{equation}
		
		In Figure 4(C), the Jeans critical number density is plotted as a function of source size for different 
                gas temperatures ($T$=10 K, 20 K, and 30 K) and is overlayed with the observed results. 
		Here, we consider that the gas temperature within OMC filaments is approximately 20 K (e.g., Cesaroni \& Wilson 1994). 
		The plotted results clearly show that most SMA sources are located very close to the critical line ($T_{\rm{gas}}$=20 K) and 
		several sources have a larger size than the Jeans length at their measured density.  
		This result suggests that most of the detected continuum sources are gravitationally unstable and could be collapsing 
                if we assume effect only from thermal support. 
		No significant differences are detected between prestellar cores and protostellar cores.

	\subsubsection{Evolutionary Stages of the Continuum Sources}
		
		The Jeans criterium indicates that most of our SMA sources are gravitationally unstable. 
		Here, we examine other observational evidence of their protostellar nature. 
		In Figure 2, we compare the positions of infrared sources, i.e., protostars (denoted by orange open circles), and 
		T Tauri stars (denoted by white open circles) based on Peterson \& Megeath (2008), and our  850 $\mu$m continuum sources; 
		these comparisons are of similar spatial resolutions (2--5$''$ corresponding to \hbox{830--2100 AU} at Orion). 
		Multi-wavelength data and outflow survey results are summarized in column 2--7 of Table 5. 
		{\it Spitzer} \hbox{8 $\mu$m} sources were detected toward eight out of twelve SMA sources 
		\hbox{(${\sim}67\%$)}, six out of twelve SMA sources \hbox{(${\sim}50\%$)} have CO bipolar outflow, 
                while three out of twelve SMA sources \hbox{($\sim$25\%)} are 
		associated with radio jets (Aso et al. 1998; Reipurth et al. 1999; Stank et al. 2002; Williams et al. 2003; 
                Takahashi et al. 2006; 2008; 2009; Takahashi \& Ho 2012). 
		These results suggest that approximately half of the SMA sources in this region are protostellar.
		
		Eight out of 16 Spitzer sources (within the SMA observing region) are associated with the SMA continuum sources, 
                which are identified as protostars (Peterson \& Megeath 2008; Megeath et al. 2012).  
		Six out of 16 Spitzer sources, which are categorized as T-Tauri stars have no counterparts in the SMA continuum emission. 
		Two out of 16 Spitzer sources identified as protostars (Peterson \& Megeath 2008; Megeath et al. 2012) are not associated with SMA continuum sources. 
                 We did not detect submillimeter continuum emission from the T Tauri source as well as two protostars likely due to the mass 
                detection limit of our SMA observations. Evolutionary status of the individual sources are summarized in the last 
                column of \hbox{Table 5}. 
		Here, the classification between prestellar and protostellar cores is based on the detection of either an infrared source, and/or 
		molecular outflow, and/or radio jet. 
		The evolutionary status of a protostellar core was determined based on its SED. Since the spatial resolution of 
                submillimeter observations are limited, 
		we could not use the original definition for identifying Class 0 sources (i.e., $L_{\rm{bol}}/L_{\rm{submm}}<200$ by Andre et al. 1993). 
		Instead of this, the classification of Class I (Class 0) sources was determined based on the detection (non-detection) of 2.2 $\mu$m emission.

	\subsection{Filament Fragmentation}	
						
	\subsubsection{Hierarchical Structure}
		Our SMA continuum observations with 4.5$''$ angular resolution have revealed a chain of density enhancements 
		within the OMC-3 filament. 
		The large-scale filamentary structure and their fragmentation properties can be a key to understanding the initial 
                conditions of embedded clusters such as core spatial distribution, mass distribution, number distribution, 
                and evolutionary sequence, all of which may govern the initial core mass 
		function of young stellar clusters and initial stellar multiplicity. 
		
		Certain fragmentation length scales within the \hbox{OMC} filaments have been suggested by several 
                previous studies (Dutrey et al. 1991; Hanawa et al. 1993; Wiseman \& Ho 1998; Johnstone \& Bally 1999). 
		The largest spacing is shown in the \hbox{$^{12}$CO} image by Maddalena et al. (1986), which corresponds to 
		the \hbox{Orion A} and \hbox{Orion B GMCs}, and has a separation of \hbox{${\sim}$4.7$^{\circ}$} \hbox{(35 pc)}. 
		Within the Orion A cloud, periodical structures have been reported by Dutrey et al. (1991) in the \hbox{C$^{18}$O (1--0)} 
		image with peak separations about \hbox{9$'$ (1.1 pc)} toward the northern part of the \hbox{Orion A} cloud, 
                while similar structures, with \hbox{${\sim}10'$} (1.2 pc) separation, have also been reported in the $^{13}$CO (1--0) 
                image at the southern part of the \hbox{Orion A} cloud by Hanawa et al. (1993). 
		The emission peaks correspond roughly to significant star-forming regions such as \hbox{OMC-1S,} \hbox{OMC-1,} 
                \hbox{OMC-2,} and \hbox{OMC-3} within the \hbox{Orion A} cloud (Johnstone \& Bally 1999). 
		Single-dish observations in the \hbox{NH$_3$} \hbox{(1, 1)} and \hbox{(2, 2)} and single-dish \hbox{(sub)millimeter} 
                continuum have shown the \hbox{0.1 pc} scale dense clumps embedded within the \hbox{OMC} filaments 
                (Cesaroni \& Wilson 1994; Chini et al. 1997; Lis et al. 1998; Johnstone \& Bally 1999). 
		Johnstone \& Bally (1999) pointed out that these clumps are equally separated along the filaments with separation of 
		\hbox{${\sim}150''$} \hbox{(0.31 pc).} Moreover, smaller size scale density enhancements have been found 
                in the high-angular resolution \hbox{VLA NH$_3$} observations of the \hbox{OMC-1} region 
		(Wiseman \& Ho 1998). Their mean separation is \hbox{${\approx}50''$} (0.1 pc). 
		
		Our SMA continuum observations revealed fragmentation within the \hbox{0.1 pc} clumps. 
                Nearest neighbor distances for each source are calculated, and listed in the sixth column of Table 4. 
                The median separation between nearest neighbors is measured to be ${\sim}17''$ or 0.035 pc
                \footnote{For the nearest neighbor calculation, we exclude SMM 7 since the separation between SMM 7 and SMM 8, 
                ${\sim}1.4''$, are much smaller than the other sources. We consider that the separation of the sources are likely 
                related to form a binary system, which is the next fragmentation size scale.}. 
                Furthermore, we have performed \hbox{1.3 mm} continuum observations with the SMA toward the \hbox{OMC -1} north region, 
                and found similar source separation within their filaments ($\approx$0.06 pc). 
                The details of the OMC-1 north observations will be presented and discussed in Teixeira et al. (2013). 
				
		In summary, our SMA observations have resolved the continuum sources with quasi-periodical separation as small as 
                \hbox{$\sim$0.035 pc}. A comparisons of the SMA continuum observations with previous single-dish observations 
		suggests \hbox{multi-spatial} scale fragmentation within the \hbox{Orion A} GMC with sizes scale 
                between \hbox{37 pc} and \hbox{0.035 pc}.

		\subsubsection{Jeans Analysis}
		
		The fragmentation of molecular clouds is related to gravitational instabilities. Jeans analysis can provide a useful 
		way to discuss which scale can grow and produce high-density regions. As described in Section 4.1.1, the critical 
                Jeans length is described as equation (4) with the assumption of an initially uniform background density 
                (e.g., Jeans 1902; Binney \& Tremaine 1987). 
		For other geometric configurations, such as cylindrical and sheetlike structures, sufficiently large perturbations can also result in 
		gravitational instability (e.g., Nakamura, Hanawa \& Nakano 1993; Burkert \& Hartman 2004). 
		In the case of an infinitely long static  and cylindrical isothermal cloud, 
		the perturbation wavelength for maximum instability (hereafter we simply call this the Jeans length) 
		is described as follows (e.g., Nakamura, Hanawa \& Nakano 1993; Wiseman \& Ho 1998):  
				
		\begin{equation}
				{\lambda_{\rm{frag.}}}{\sim}{\frac{20c_s}{{\sqrt{4{\pi}G{{\rho}_0}}}}}.
		\end{equation}

		In order to understand the physical mechanism of fragmentation processes over a wide range of size scale, 
		the measured source separation is plotted as a function of the number density of their respective parental 
                molecular clouds/clumps/cores in \hbox{Figure 5}. 
                Note that the inclination of the filaments should be taken into account in this discussion. 
		According to Genzel \& Stutzki (1989), a randomly oriented filament will have a line of sight mean (median) 
                inclination angle of 45$^{\circ}$ (60$^{\circ}$). 
		To display this projection effect, \hbox{(${\lambda}_{\rm{obs}}={\lambda}_{\rm{frag}}{\cdot}{\sin}i$)}, 
		core separations in the case of \hbox{$i=45^{\circ}$} are plotted in \hbox{Figure 5} as open symbols. 
		Table 6 summaries the number densities, structure separations, and gas temperatures of different structures. 
		These structures are listed by size: GMCs ($\geq$1pc), large-scale clumps (0.5--1.0 pc), small-scale clumps 
                (0.1--0.5 pc), and dense core ($\leq$0.1 pc). 
		Here, we assume that dense cores are formed via the fragmentation of parental small-scale clumps, 
                which in turn are formed via the fragmentation of parental 
		large-scale clumps, and that these latter structures are formed via the fragmentation of parental GMCs. 
		The measured separations were plotted as a function of the number density of the parental large scale structure.
		To further discuss the smaller scale fragmentation properties, the separation of the visible binary stars studied in 
		the Orion nebula cluster (ONC; Reipurth et al. 2007) are also plotted assuming that 
		the number density of their parental core is similar to that of our SMA continuum source. 
		
		We overplot on Figure 5 the perturbation wavelength, $\lambda_{\rm{frag.}}$, given in equation (2) and (4) 
                for various gas temperatures. The observed fragmentation spacings (denoted in Figure 5 as filled symbols) 
                are roughly following a single power law over almost four orders of magnitude, suggesting that the hierarchical 
                fragmentation could be governed by a single physical process. 
		Figure 5 indicates that the data is described better by a shallower index, -1.4 (dashed line), 
                than that predicted maximum instability size as a function of density of an isothermal clouds, -2.0 (solid line). 
		This shallower index can be explained by differences of cloud temperatures over different size scales. 
		Parsec-scale GMCs are exposed to ionized gas from HII regions, cosmic rays, large-scale turbulence, etc., 
                which work as heating processes. Consequently, the mean temperature of parsec scale clouds tends to be 
                relatively high (e.g., \hbox{50-100 K} in the ISMs; e.g., Ferriere 2001).  
		Intermediate regions (both large- and small-scale clumps) of the OMC molecular cloud (0.1-1.0 pc) is also most likely exposed to the strong 
		interstellar radiation field from nearby star forming clusters, 
		e.g., strong heating from the exterior, corresponding to 1000 times stronger interstellar radiation field as suggested 
		toward a few small-scale clumps in the OMC-2/3 region from the CO isotopes observations 
                with Monte Carlo line radiative transfer models (e.g., J$\o$rgensen et al. 2006). 
		This strong radiation is able to produce the higher gas temperature ($>$30 K) at the outer most part of the envelope gas. 
		Finally, deeply embedded dense cores within filaments are shielded from these heating processes. Thus, the innermost parts of clouds have 
		low temperatures (e.g., \hbox{${\approx}20$ K}; Cesaroni \& Wilson 1994; Wiseman \& Ho 1998; Tatematsu et al. 2008). 
		This decrease in temperature with size scale, means the fragmentation length will also decrease more sharply than the isothermal case. 
		This leads to a more shallow power law in Figure 5. 
		
		Moreover, the dissipation of the magnetic flux on the smaller scale can also be responsible for the observed shallower power law index. 
		Since large-scale diffuse structures are always exposed to cosmic ray and/or ionized gas from the surrounding HII region, 
		the fractional ionization in large-scale molecular clouds is higher than in the small scale dense gas core. 
		Therefore, ambipolar diffusion will be more efficient in the dense cores, leading to less 
		magnetic support, and a smaller fragmentation length (e.g., Nakano \& Tademaru 1972). 
		Finally, turbulence, which is often observed in GMC size scales (Larson 1981; Solomon et al. 1987), can also be responsible for the observed 
                shallower power law index. Detailed discussion for the OMC case will be described in the Section 4.2.3. 
 
		Observational results suggest that the fragmentation spacings, especially at the scales of clumps (\hbox{0.1--1 pc}), are roughly 
		consistent with the maximum instability size derived from either the initially uniform density cloud 
                (orange solid line in Figure 5) or cylindrical clouds (blue solid line in Figure 5) assuming increasing 
                temperatures with size scale.  
		This indicates that the fragmentation may be occurring through the thermal fragmentation process. 
		The SMA detected continuum sources (denoted in Figure 5 by triangle shaped data points) are embedded within a filament. 
		Hence the core separation should be around the maximum instability size of the cylindrical clouds model 
                (denoted in Figure 5 by blue line). However, the observed core separation is a factor of 2.0--7.3 smaller than the 
		thermal fragmentation spacing with \hbox{$T_{\rm{gas}}$=20 K}, $n_{\rm{H_2}}=8.8{\times}10^4 - 1.1{\times}10^6$ cm$^{-3}$ (Table 6), 
                and $i$=90$^{\circ}$. In the next subsection, we explore different physical processes that can affect the fragmentation size scale.

		\subsubsection{What Mechanisms affect the Observed Core Separation?}
		From our SMA observational result, the observed fragmentation ($\lambda_{\rm{frag}}$=0.035 pc) is smaller 
                than the Jeans length of the clump. Projection effects are not sufficient to explain this difference: 
                $\lambda_{\rm{frag}}=\lambda_{\rm{obs}}/{\sin}i<{\lambda_J}$ for $45^{\circ}{\geq}i{\geq}90^{\circ}$. 
		In order to further discuss the fragmentation properties within the Orion A GMC, we summarize different possible mechanisms that can 
		affect the fragmentation size scale.  
		
		{\bf Turbulence:} 
		In this case, the cloud fragmentation length is described by equations (2) and (4) with the effective sound speed 
                \hbox{$c_{\rm{eff}}$} instead of the sound speed \hbox{$c_s$}. 
		The $c_{\rm{eff}}$ is defined as $c_{\rm{eff}}^2={\frac{{\Delta}{v_{\rm{obs}^2}}}{8{\ln}2}}$, 
                where ${\Delta}v_{\rm{obs}}$ is the observed velocity dispersion 
		and is given by $\Delta{v_{\rm{obs}}^2}={\Delta{v_{t}^2}+{\Delta}v_{nt}^2}$, 
		where $\Delta{v_{\rm{t}}}$ and $\Delta{v_{nt}}$ correspond 
		to the thermal and non-thermal velocity dispersion, respectively. 
		Turbulence will therefore produce a fragmentation scale larger than the Jeans length. 
		Figure 5 and Table 6 present that observed clump separations are roughly consistent, or smaller, 
                than the thermal fragmentation lengths for scales of a few pc. 
		These results suggest that turbulence has already decayed for size scales $\leq$1 pc. 
		For scales of a few pc or larger, turbulent pressure may in fact be dominant. 
		For example, Jackson et al. (2010) studied a filament that spans 80 pc and conclude that its observed 
                fragmentation scale ($\approx$5 pc) can only be described by an isothermal filamentary cloud determined 
                by turbulent pressure, or equivalently, by a self-gravitating incompressible fluid cylinder. 
		Our data indicates that the OMC 3 filament fragmentation cannot be explained by this scenario. 
		
		{\bf Magnetic field and cloud rotation:} 
		Other physical processes that might affect the fragmentation length of filaments are helical magnetic 
                field and cloud rotation (Fiege \& Pudritz 2000; Matsumoto, Nakamura, \& Hanawa 1994). In fact, 
                Matthews et al. (2001) use helical field geometry of magnetic field to interpret their 
		JCMT/SCUBA polarization observations toward OMC-3. 
		Their data indicate that the magnetic field in the northern part of OMC-3 is toroidal. 
		Hanawa et al. (1993) has performed magnetohydorodynamical simulations for filament fragmentation, 
		consisting of an isothermal rotating cloud + helical magnetic field, and found  
		a dispersion relation (eq 11, Hanawa et al. 1993). We can write the fragmentation wavelength 
                for the toroidal magnetic field case as: 
				
		\begin{equation}
			\lambda_{\rm{frag}} = \frac{2{\pi}H}{0.72[(1+{{\alpha}/2}+{\beta})^{1/3}-0.6]}
		\end{equation}
		 
		 \noindent
		 where $H$ represents a length scale of the parental clump, and given by 
		 
		\begin{equation}
			H^2 = \frac{c_s^2}{4{\pi}G{\rho_c}} \left( 1+\frac{1}{2}{\alpha}+{\beta}  \right)
		\end{equation}

		\noindent 
		where $\rho_c$ is the peak density of the parental clump, and the effect from magnetic field and rotation 
                are described by $\alpha$ and $\beta$ as: 
		 
		 \begin{equation}
		 	\alpha = \frac{B^2}{8{\pi}{\rho}_c{c_s^2}},~~~{\beta} = \frac{2{\Omega}^2H^2}{c_s^2}
		 \end{equation}		
		
		\noindent
		$\Omega$ is the rate of rotation and $B$ is the magnetic field strength.  
		Equation (5) shows that if \hbox{(${\alpha}+{\beta}$)}$\geq$1, 
		the maximum instability size will become smaller than the thermal fragmentation length. 
		If \hbox{${\alpha}+{\beta}<<1$}, the effects from the magnetic field and rotation are negligible.

		{\bf Global collapse of the parental cloud} 
		Recent theoretical studies present how cloud structures and global collapse significantly affect 
                local fragmentation properties (e.g., Burket \& Hartmann 2004; Pon et al. 2011; Pon et al. 2012). 
		If the original parental cloud is not infinitely long or spherical, gravitational focusing results 
                in enhanced concentrations of mass inducing global collapse 
		unless magnetic fields, cloud rotation, and/or external pressure are dominated (Burket \& Hartmann 2004).
		As we can see from the JCMT/SCUBA 850 $\mu$m continuum observations (figure 2, right panel), 
		the parental filamentary clump (e.g., 1.5 Jy beam$^{-1}$ denoted by thick contour) shows a relatively well defined edge, 
		so according to Burket  \& Hartmann's model, global collapse could potentially be underway in this clump. 
		A recent paper by Pon et al (2011) pointed out that filamentary geometry is the most favorable situation 
                for the smallest perturbation to grow 
		before global collapse overwhelms the cloud. Their following paper (Pon et al. 2012) suggested that filaments 
                with large aspect ratio would take more time to collapse. For example, the global collapse  time-scale for 
                a cloud aspect ratio of 10 (similar case of the observed clumps) 
		is roughly seven times slower than the spherical free-fall time. 
		This can explain how local fragmentation at the size scale of dense cores and global collapse of their 
                parental small-scale clump could be concomitant. 
		Global collapse will eventually shorten the spacing between the fragmented cores, and this scenario could 
                therefore potentially explain the observed shorter fragmentation length compared to Jeans length. 
		
		In summary, there are three physical mechanisms that can increase (turbulence) or decrease 
                (magnetic fields and rotation, or global collapse) the fragmentation length relative to the Jeans length. 
                The fragmentation lengths obtained from the data presented in this paper are consistent (0.3 pc -1pc scale) 
                or smaller ($\leq$ 0.1pc scale) than the Jeans fragmentation length, suggesting that turbulence does not play 
                a significant role in the fragmentation process for these size scales. 
                Note that some of their core are larger than their associated Jeans length, therefor, they could fragment further. 
                We discussed how the magnetic field and cloud rotation, or global collapse may decrease the observed fragmentation length. 
                However, at the moment we are unable to quantitatively constrain $\alpha + \beta$, or inflow from the global 
                collapse for our case -- only future direct comparisons between higher-dynamic range kinematic data and more 
                realistic models can properly address these issues.

		{\bf Core Fragmentation:} 
		In addition to the clump large-scale fragmentation properties discussed above, we would like to explore 
                further the smaller core-scale fragmentation properties, 
		which is related to binary star formation ($\leq$1000 AU). 
		These binary stars observed in the Orion Nebula Cluster (ONC) 
		have a wide range of separations from \hbox{0.0027--0.027 pc} (\hbox{67.5--675 AU}; Reipurth et al. 2007). 
		The plotted results in Figure 5 suggest that wide binary systems ($\geq$200 AU) might follow the same power 
                law index extrapolated from the observed large size scale cloud fragmentation, i.e., these wide binaries 
                could be formed through the next level of hierarchical process.  
		This lends support to the theoretically predicted binary formation scenario through 
		the hierarchical fragmentation of the parental core occurs in the isothermal collapse phase 
                (e.g., Goodwin et al. 2007; Machida et al. 2008). 
		Close binary systems ($\leq$100 AU) might be formed through other mechanisms; 
		hydrodynamic simulations suggest that angular momentum within the rotating and collapsing hydrostatic 
                first cores is a crucial parameter to determine the later fragmentation properties 
                (e.g., Boss 1989; Bonnell and Bate 1994; Bate 1998; Saigo \& Tomisake 2006; Machida et al. 2008). 
		Furthermore, star-disk encounters could also play an important role in re-distributing angular momentum 
                in protoplanetary disks and forming binary systems (e.g., Larson 2002; Pfalzner 2004; 2005).

	\subsubsection{Fragmentation Timescale}		
		One interesting issue related to filament fragmentation is whether the fragmentation occurs simultaneously 
                in the entire filament, or instead propagates along it producing sequential star formation along the filament. 
		With the SMA continuum observations, we found the following interesting tendencies among the sources; 
		(i) a chain of continuum sources is detected in the northern part of the \hbox{OMC-3} filament from 
                \hbox{SMM 2} to \hbox{SMM 8} with a similar spatial separation, 
		(ii) a relatively extended fluffy source (elongated structure), \hbox{SMM 5}, is located at the center 
                of the chain of the continuum sources, and this fluffy structure seems to be still in the process of core 
                formation (or fragmentation). 
		No star formation signatures such as outflow/jet association or infrared source are found, 
		which supports the idea that \hbox{SMM 5} is probably the youngest star formation site within the filament. 
		Why does \hbox{SMM 5}, which is located in the middle of the filament, have the fluffy structure? 
		A fragmentation process which propagates along the filament from each end inwards can  be one possibility to explain this. 
		
		Previous numerical calculations give us a hint that the finite geometry of gas clouds promotes concentrations of 
		material at the ends of the filament, at the beginning of the filament fragmentation. 
		Hanawa, Yamamoto, \& Hirahara (1994) performed one-dimensional hydrodynamic simulations of a finite long 
                filamentary cloud. These simulations suggest that dense cores are produced with a semi regular interval in space 
                and time from one edge to the other. The gas near the edge is attracted inwards by gravity and the accumulation 
                of the gas makes a dense core near the edge. 
		When the dense core grows in mass up to a certain amount, it gathers gas from the other direction. 
		Accordingly the dense core becomes isolated from the main cloud and the parent filamentary cloud has a new edge. 
		Via the same process, the next accumulation of gas at the new edge makes a new dense core. 
		The fragmentation is considered to propagate with approximately the effective sound speed (\hbox{1.23$c_{\rm{eff}}$} 
                derived from the simulation by Hanawa Yamamoto \& 
		Hirahara 1994; \hbox{1.1$c_{\rm{eff}}$} derived from a linear stability analysis  by Nakamura, Hanawa, \& Nakano 1993). 
		Signatures of the fragmentation propagating down the filament is actually observed in the Taurus Molecular Cloud -1 
                region (Hanawa, Yamamoto \& Hirahara 1994). 
		
		Our SMA results also show that the chain of continuum sources, from \hbox{SMM 2} to \hbox{SMM 8}, are embedded 
                within the most northern part of 
		the \hbox{OMC-3} filaments where both ends show edge-like density contrast (finite filamentary clouds with two edges; 
                \hbox{Figure 2}). The mean value of the observed velocity width in \hbox{FWHM} is approximately \hbox{0.88 km s$^{-1}$} 
                in this region (A mean velocity width derived from NH$_3$ observations toward the norther OMC 
                cloud by Cesaroni \& Wilson 1994) and the effective sound speed is 
		estimated to be \hbox{$c_{\rm{eff}}{\sim}\sqrt{\frac{{\Delta}v^2}{8 ln2}}{\sim}$0.37 km s$^{-1}$}. 
		The mean projected distance from the sources at the two edges of the filament (i.e., \hbox{SMM 2} and \hbox{SMM 8}) 
                to \hbox{SMM 5} is measured to be \hbox{0.15 pc}. 		
		Therefore, the time scale of the fragmentation propagation is estimated to be approximately 
                \hbox{0.15 pc/0.37 km s$^{-1}{\sim}$3.0$\times$10$^5$ yr}. 
		The detected continuum source in the filament have similar to the protostellar phase (Peterson \& Megeath 2008; 
                Megeath et al. 2012). The typical evolutionary time scale of such sources are \hbox{10$^4$--10$^5$ yr} (Beichman et al. 1986; Shu et al. 1977). 
                Estimated propagation time scale of the filament fragmentation is 
		similar to, or slightly longer than, the star formation time-scales of the detected sources. 
		This may suggest that the protostars formed within the fragmented cores, where star formation began almost at the same time as the 
		clump fragmentation. More complete fragmentation time scale analysis (including other parts of the OMC filaments) 
                will be presented in following papers.

\section{CONCLUSION \& FUTURE PROSPECTS}
	High-resolution \hbox{850 $\mu$m} continuum observations have been carried out using the SMA toward the \hbox{OMC-3} northern filament with 
	a mosaic of \hbox{85 fields} covering approximately \hbox{$6'.5{\times}2'.0$} (\hbox{0.80$\times$0.25 pc}). 
	The data were supplemented by multi-wavelength data sets for investigating evolutionary stage of each source. 
	The main results are summarized as follows:

\begin{enumerate}

	\item	 With the SMA \hbox{850 $\mu$m} observations, 12 dust continuum sources have been identified and spatially 
        resolved with an \hbox{H$_2$} mass between \hbox{0.3 --5.7 M$_{\odot}$} and the source size between \hbox{1400-- 8200 AU}. 
	We found 850 $\mu$m continuum counterpart to the previously known ten sources and present the discovery of two new additional sources. 
	The detected sources have a variety of structures and evolutionary stages, ranging from prestellar phase to the Class 0/I phases. 
	The Jeans analysis suggests that all the detected sources are most likely gravitationally unstable. 
	Multi-wavelength data sets confirmed that 6/12 (50 \%) continuum sources have molecular outflows, 8/12 (67 \%) continuum sources are 
	associated with infrared sources, and 3/12 (25 \%) continuum sources are associated with ionized jets.  
	These results suggest that the OMC-3 is an active protostar forming region. 
		
	\item	 Hierarchical structure with quasi-periodical separations on different size-scales is suggested 
        with a size of \hbox{${\sim}4.7^{\circ}$/35 pc} down to \hbox{$17''$/0.035 pc}, 
	implying a multi-size scale fragmentation process. 
	Fragmentation spacings in clumps \hbox{(0.1--1 pc)} are roughly consistent with the thermal fragmentation lengths in each structure. 
	However, shorter spacings than those expected from thermal fragmentation have been detected in the SMA data ($\approx$0.04 pc). 
	This could be explained by either an helical magnetic field, parental cloud rotation, or/and large-scale filament collapse. 
		
	\item	 Possible evidence for sequential fragmentation as local conditions evolve is suggested in the northern part 
        of the \hbox{OMC-3} filament. The fragmentation propagation timescale is estimated to be  
        \hbox{$\tau_{\rm{frag.}}{\sim}3.0{\times}10^5$ yr}. This time scale is roughly consistent with (or slightly longer than) 
        the life time of the detected protostellar sources, which is \hbox{10$^4$--10$^5$ yr}. 
	These results suggest that the protostars formed within the fragmented cores, and that star formation began at almost 
        the same time as the core fragmentation.  
	
\end{enumerate}

Our interferometric observations successfully resolved the detailed gas structure associated with the individual star formation sites 
within the filamentary structure. However, present observations are still limited by mass sensitivity ($\geq$0.15 M$_{\odot}$) and 
suffer from the large missing flux of the filament (${\sim}$16\% recovering flux). Due to this, we are biased to detect relatively 
bright and compact structures (i.e., mostly massive envelope associated with the pre-stellar and protostellar cores). 
In order to detect less bright sources (e.g., emission from Class II sources, brown-dwarf mass sources) 
and to obtain a wider evolutionary range from the sample, even higher sensitivity observations are required. 
In addition, a larger sample is crucial to investigate the evolutionary tendencies along the entire OMC filaments.
Furthermore, polarization observations with high sensitivity and high-spatial resolutions are crucial for elucidating the magnetic field structures, 
for measuring the field strength at the early evolutionary stages, and for verifying the role of the magnetic field in the filament fragmentation processes.
These issues are future prospects with extended data sets using the SMA, EVLA, and ALMA.

\acknowledgments
We acknowledge the staff at the Submillimeter Array and the Very Large Array for assistance with operations and helpful comments 
related to data reduction. S. T. acknowledges K. Asada, B.-C. Hsieh, Z.-W. Zhang, A. Hacar, K. Saigo, H. Hirashita, D. Johnstone for fruitful comments. 
We also thank D. Johnstone for providing us the submillimeter continuum data taken with the JCMT/SCUBA.
The authors are grateful to the referee for the constructive comments that have helped to improve this manuscript. 
S. T. is financially supported by a postdoctoral fellowship at the Institute of Astronomy and Astrophysics, Academia Sinica, Taiwan, 
and was supported by visitor programs at ESO (Garching) and MPIfR, whilst visiting co-authurs P.S. Teixeira and L. A. Zapata, respectively.

\clearpage

\begin{figure}
\rotate
\epsscale{0.8}
\plotone{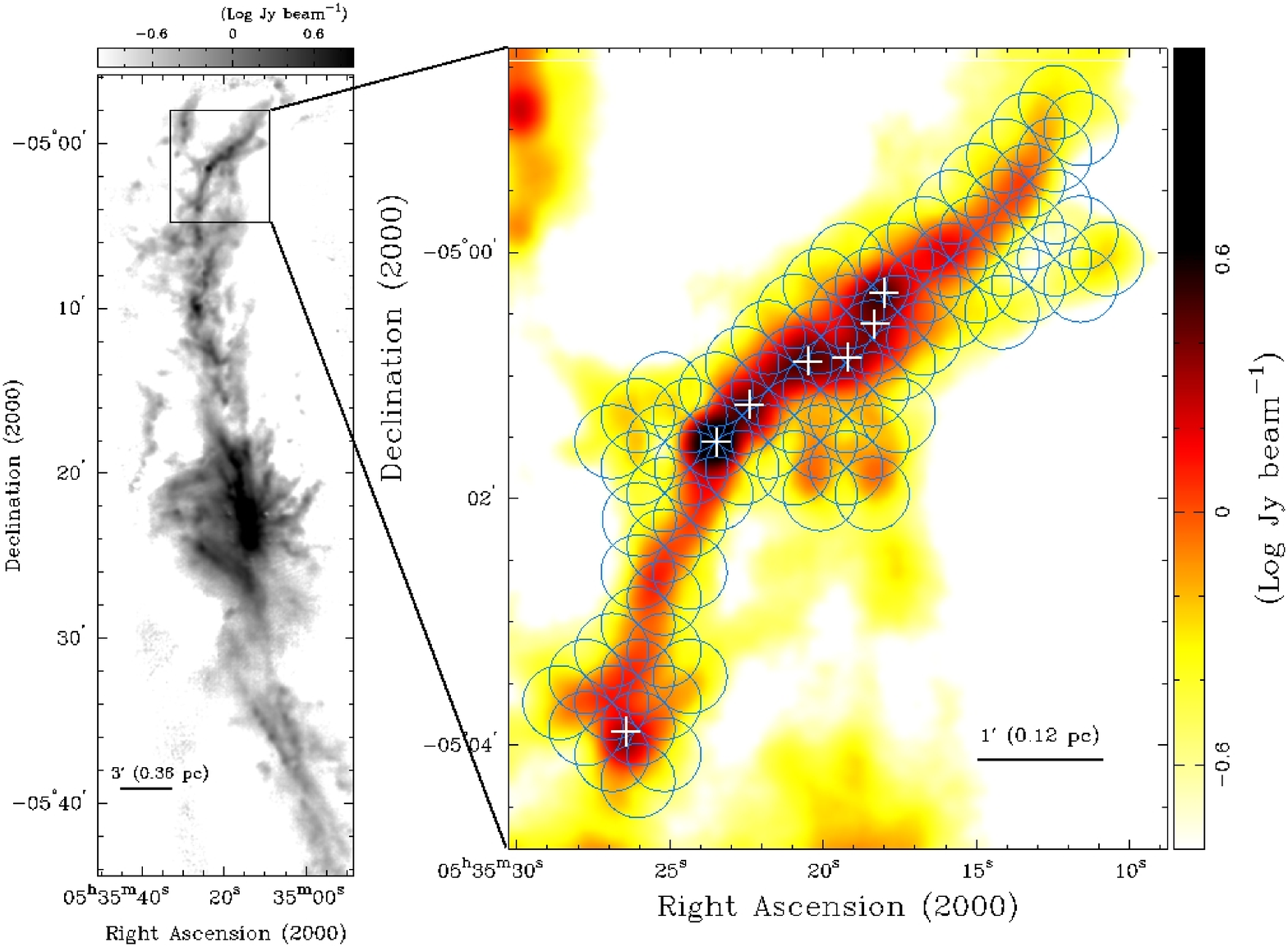}
\caption{Left panel: An \hbox{850 $\mu$m} dust continuum image taken with the \hbox{JCMT/SCUBA}, referred from \hbox{Johnstone \& Bally (1999)}. 
Right panel: SMA 85 points mosaic. The background image is same as in the left panel. Open circles show the SMA primary beam at \hbox{351 GHz}. 
Crosses in the image show the positions of previously detected \hbox{1.3 mm} continuum sources with a single-dish telescope (Chini et al. 1997).  
\label{f1}}
\end{figure}
\clearpage

\begin{figure}
\rotate
\epsscale{1.1}
\plotone{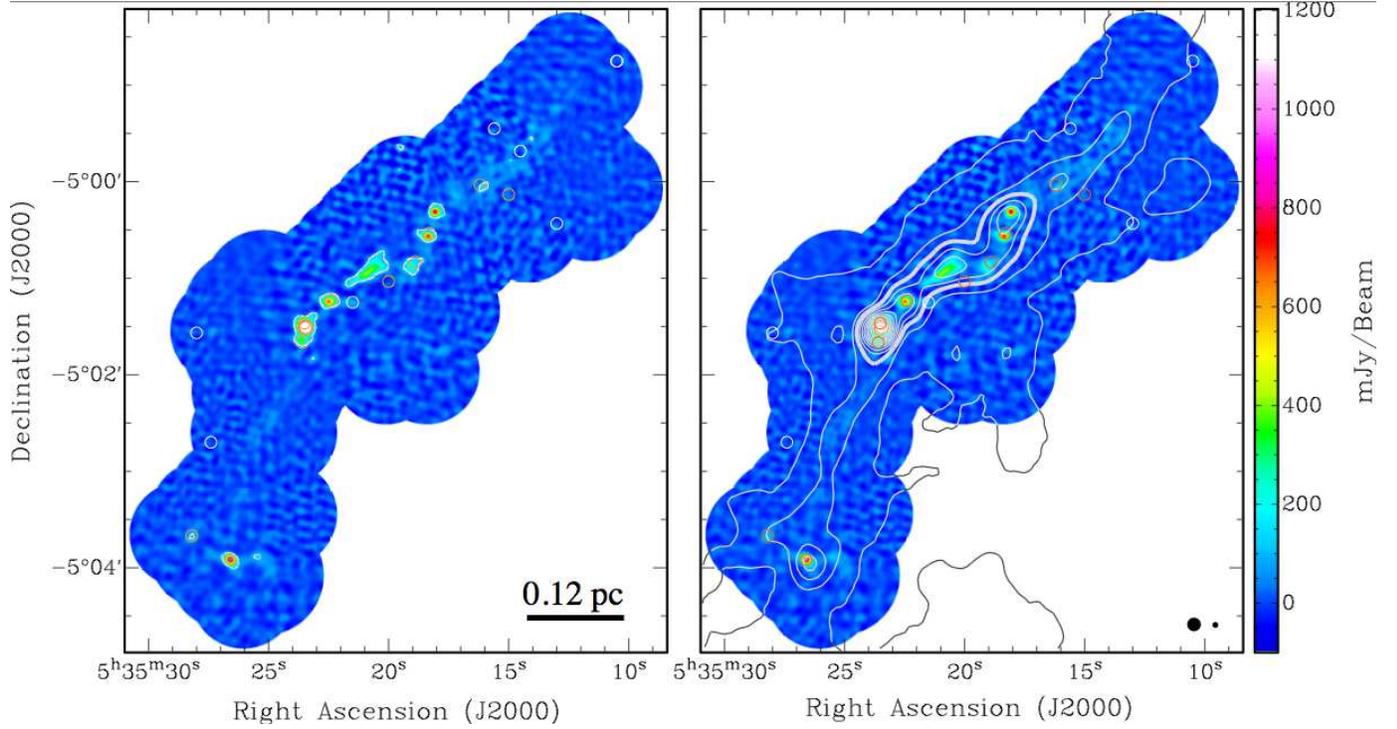}
\caption{Left panel: An \hbox{850 $\mu$m} continuum mosaic image taken with the SMA. 
White contours show the \hbox{5$\sigma$} signal levels (i.e., \hbox{100 mJy beam$^{-1}$}). 
Orange open circles and white open circles present the positions of protostars and T Tauri stars, respectively \hbox{(Peterson \& Megeath 2008)}.
Right panel: The SMA \hbox{850 $\mu$m} continuum mosaic image (color) overlaid 
with the \hbox{850 $\mu$m} continuum image taken with the \hbox{JCMT/SCUBA} (contours referred from \hbox{Johnstone \& Bally 1999}). 
The contour levels start at \hbox{0.3 Jy beam$^{-1}$} with a interval of \hbox{0.6 Jy beam$^{-1}$}. 
The orange and white circles are the same as in the left panel. 
Black filled circles in the bottom right corner show the JCMT and SMA beam sizes of 14$''$ and $4''.5$, respectively. 
Thick contour (contour level of 1.5 Jy beam$^{-1}$) presents a relatively well defined edge of the small-scale clump discussed in Section 4.2.3.
\label{f1}}
\end{figure}
\clearpage

\begin{figure}
\rotate
\epsscale{1.0}
\plotone{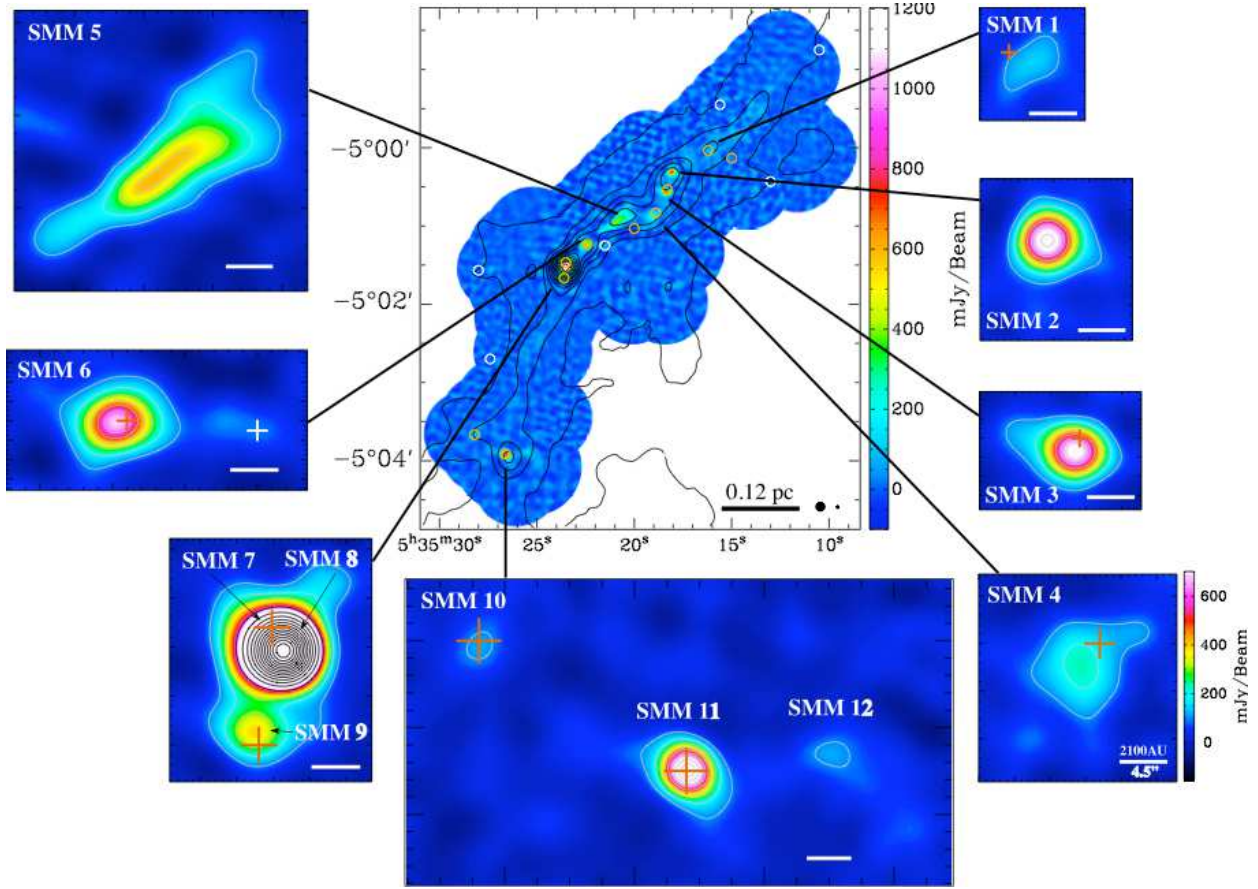}
\caption{Zoomed-in images of SMA detected continuum sources. The contour levels of these images (except the \hbox{SMM 7--9} panel) 
start at \hbox{5$\sigma$} level (i.e., \hbox{100 mJy beam$^{-1}$}) with a interval of \hbox{5$\sigma$}. 
Contour levels of \hbox{SMM 7--9} show \hbox{5, 10, 15, 30, 45, 60, ... 210 $\sigma$}. 
In these data SMM 7 is not resolved, Takahashi et al. (2009) shows the source resolved and accurately positioned. 
Positions of \hbox{8$\mu$m} {\it Spitzer} sources identified by \hbox{Peterson \& Megeath (2008)} are denoted by orange crosses. 
The bars in the bottom right corner in each panel show the scale of \hbox{2100 AU}, which is the SMA beam size in FWHM. 
Color wedge denoted at the SMM 4 panel shows the signal levels of all the individual panels.
The filled circles in bottom right corner of the central panel presents the SMA and JCMT beam sizes of $4''.5$ and $14''$, respectively.
\label{f1}}
\end{figure}
\clearpage

\begin{figure}
\epsscale{1.1}
\plotone{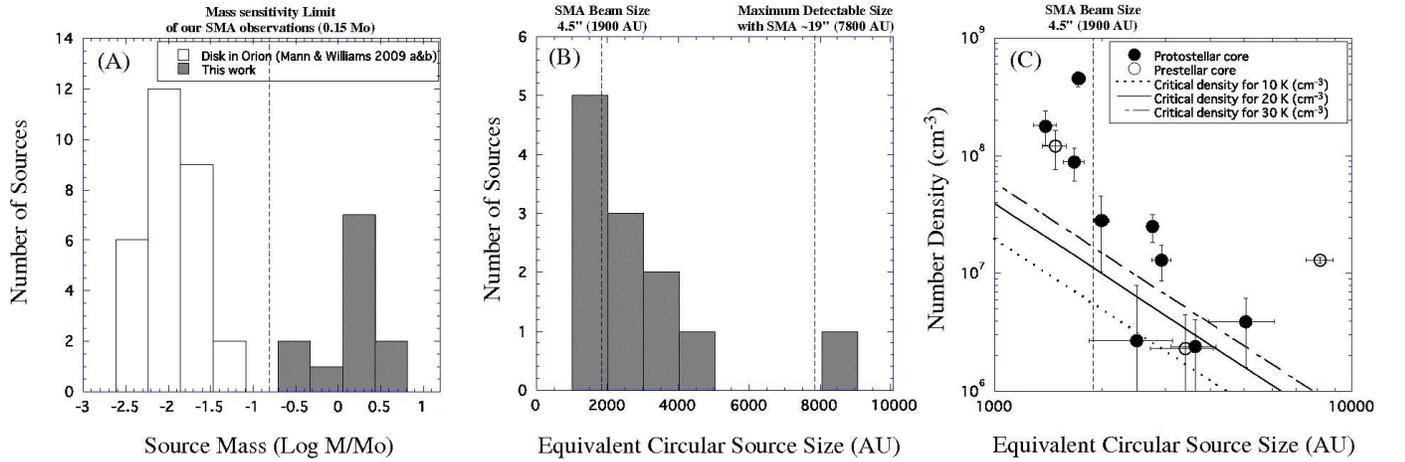}
\caption{(A) Histogram of source mass. The SMA detection limit of 1.5 M${_{\odot}}$ was estimated from equation 1 with the following assumptions: 
total flux of 100 mJy (5 $\sigma$), $T_{\rm{dust}}$=20 K,and $\beta$=1.6. (B) Histogram of equivalent circular source size, =\hbox{$\sqrt{D^2_{\rm{maj.}}+D^2_{\rm{min}}}$}, 
where $D_{\rm{maj.}}$ ($D_{\rm{min.}}$) is size of the major (minor) axis of the source in AU. 
(C) Filled and open circles show the source sizes as a function of number density for protostellar and prestellar sources. 
Dotted, solid, and dashed lines show the critical Jeans length as a function of number density with the gas temperature of $T_{\rm{gas}}$=10, 20, and 30 K, respectively.
\label{}} 
\end{figure}
\clearpage

\begin{figure}
\rotate
\epsscale{0.7}
\plotone{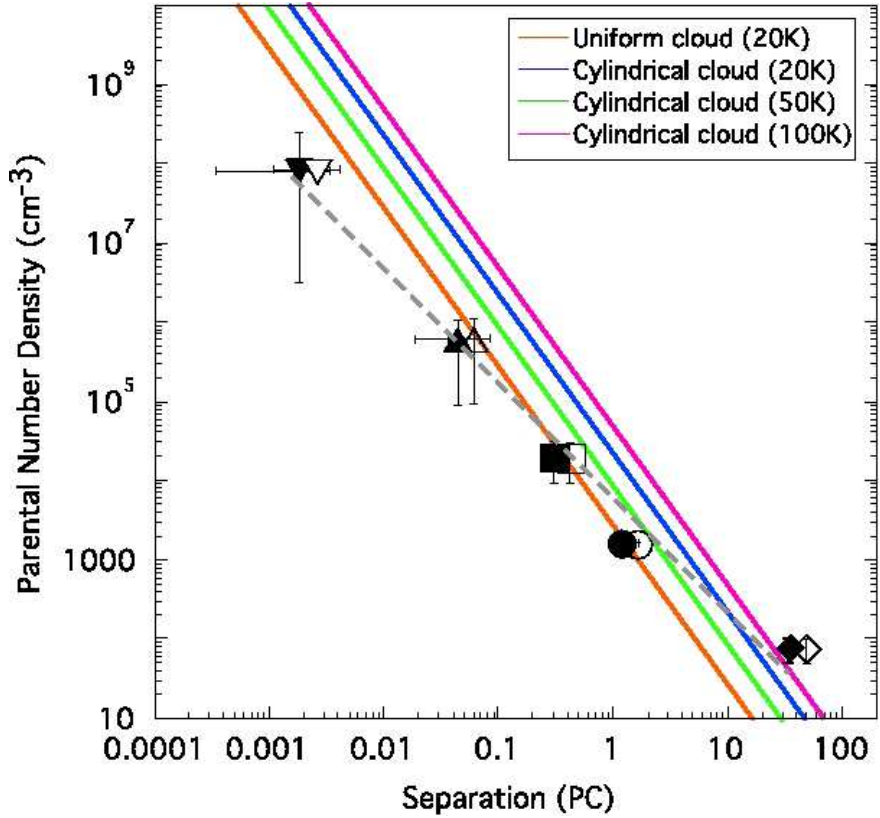}
\caption{Observed separations of clouds/large-scale clumps/small-scale clumps/cores in parsec as a function of the mean hydrogen number density of the parental cloud in cm$^{-3}$. 
The filled symbols show the separations with an assumption of $i=90^{\circ}$ (filament in the plane of the sky), and the open symbols show the separation with an 
assumption of $i=45^{\circ}$. 
The plotted values are summarized in second and fourth columns of Table 6 with reference papers. 
Data from GMC (Sakamoto et al. 1994; Wilson et al. 2005), large-scale clumps (Dutrey et al. 1991; 1993; Hanawa et al. 1993; Johnstone \& Bally 1999), 
small-scale clumps (Cesaroni \& Wilson 1994; Johnstone \& Bally 1999), dense cores (This work), and binaries (Reipurth et al. 2007) are denoted by diamonds, 
circles, squares, upright triangles, and inverted triangles, respectively. 
The dashed line shows a reference power law index of  \hbox{($n_{\rm{H_2}}{\propto}{\lambda^{-1.4}}$)}. 
The color lines show the expected maximum instability size as a function of number density derived from equations (4) and (5): 
uniform background density with a gas temperature of 20 K (orange), 
cylindrical cloud with a gas temperature of 20 K (blue line), 50K (green), and 100 K (pink). 
\label{}}
\end{figure}
\clearpage

\begin{deluxetable}{lccccccc}
\tabletypesize{\scriptsize}
\tablecaption{SMA Observational Logs\label{tbl2}}
\tablewidth{0pt}
\tablehead{
\colhead{Date} & \colhead{Configuration}  & \colhead{Number of Antennas} & \colhead{$\nu_c$(USB)/$\nu_c$(LSB)} & \colhead{${\tau}_{225}$} & \colhead{$T_{\rm{sys[DSB]}}$} & \colhead{$S_{\rm{351GHz}}^{\rm{J0423-013}}$}  & \colhead{$S_{\rm{351GHz}}^{\rm{J0530+135}}$}  \\
\colhead{} & \colhead{} & \colhead{} & \colhead{(GHz)} & \colhead{} & \colhead{(K)} & \colhead{(Jy)} & \colhead{(Jy)}  \\
}
\startdata
2008	 Jun 2        & compact          & 7  &  356.47/346.47 & 0.04--0.1  & 120--460  & 1.61  & 1.58 \\
2008 Aug 	29  	& subcompact    & 8	&  356.55/346.55 & 0.04	    &  100--200  & 1.05 &  1.43 \\
2008 Oct 21	& compact          &  8	&  356.47/346.47 & 0.08--0.1  &	240--450	& 1.06 &  1.82 \\
\enddata
\end{deluxetable}

\begin{deluxetable}{lc}
\tabletypesize{\scriptsize}
\tablecaption{SMA Continuum Observational Parameters\label{tbl1}}
\tablewidth{0pt}
\tablehead{
\colhead{Parameter} & \colhead{Value} \\
}
\startdata
Reference position (J2000.0)			                            &  $\alpha$=5$^h$35$^m$23.48$^s$, $\delta$=-5$^{\circ}$01'32.20'' \\
Configurations	  					                            &  2$\times$compact and 1$\times$subcompact \\
Primary beam HPBW (arcsec)	  			                  &  36 \\
Grid spacing (arcsec)							& 18 (P.A. 45$^{\circ}$) \\
Synthesized Beam HPBW (arcsec)	  	                           &  4.5$\times$4.5  \\ 
Equivalent frequency (GHz)                                                    & 351 \\
Total continuum bandwidth USB+LSB (GHz)	                  &  3.2 \\ 
Projected base line range (k$\lambda$)                               &  9.7--88 \\
Maximum detectable structure (arcsec)\tablenotemark{a}                                 &  17 \\
Gain calibrators                                                                          &  0423-013 and 0530+135 \\
Bandpass calibrator	                                                                 &  3C 273 and 3C 279 \\
Flux calibrators                                                             &  Titan and Neptune \\  
RMS noise level (mJy beam$^{-1}$)                                     &  $\sim$20 \\
\enddata
\tablenotetext{a}{Our observations were insensitive to more extended emission than this size-scale structure at the 10\% level (Wilner \& Welch 1994).}
\end{deluxetable}

\begin{deluxetable}{lcccccccc}
\rotate
\tabletypesize{\scriptsize}
\tablecaption{SMA 850 $\mu$m Continuum Sources\label{tbl3}}
\tablewidth{0pt}
\tablehead{
\colhead{ID} & \colhead{R.A.\tablenotemark{a}}  & \colhead{Decl.\tablenotemark{a}} & \colhead{Total Flux} & \colhead{Peak Flux\tablenotemark{f}} & \colhead{Deconvolved Size} & \colhead{P.A.} &  \colhead{$C_{\rm{(SMA/JCMT)}}$\tablenotemark{b}} & \colhead{Other mm/submm source names\tablenotemark{c}}  \\
\colhead{} & \colhead{(J2000.0)} & \colhead{(J2000.0)} & \colhead{(mJy)} & \colhead{(mJy beam$^{-1}$)} & \colhead{(arcsec)} &\colhead{(degree)} & \colhead{}  \\
}
\startdata
OMC3-SMM 1	& 5 35 16.03 & -5 00 03.07  &  354$\pm$63      & 148$\pm$19     & 8.5$\pm$1.0$\times$2.3$\pm$1.1       	& 125$\pm$5     & 0.23     & CSO 3       \\
OMC3-SMM 2	& 5 35 18.03 & -5 00 17.77  &  1070$\pm$42     & 811$\pm$20     & 2.8$\pm$0.2$\times$2.2$\pm$0.2     	& 107$\pm$13    & 0.33     & MMS 1, CSO 5\\
OMC3-SMM 3	& 5 35 18.30 & -5 00 33.01  &  1001$\pm$43     & 736$\pm$20     & 3.9$\pm$0.2$\times$1.0$\pm$0.4     	& 81$\pm$3      & 0.36     & MMS 2, CSO 6\\
OMC3-SMM 4	& 5 35 18.98 & -5 00 51.63  &  1009$\pm$52     & 220$\pm$21     & 9.5$\pm$1.9$\times$7.6$\pm$1.7       	& -36$\pm$38    & 0.31     & MMS 3, CSO 7\\
OMC3-SMM 5	& 5 35 20.88 & -5 00 56.25  &  2572$\pm$65     & 393$\pm$30     & 19$\pm$1.6$\times$5.2$\pm$0.6       	& -55$\pm$3     & 0.49     & MMS 4, CSO 8\\
OMC3-SMM 6	& 5 35 22.47 & -5 01 14.43  &  1261$\pm$47     & 810$\pm$20     & 3.9$\pm$0.2$\times$2.8$\pm$0.2     	& 100$\pm$6     & 0.37     & MMS 5, CSO 9\\
OMC3-SMM 7	& 5 35 23.50 & -5 01 29.58  & 1600\tablenotemark{d}   & 320\tablenotemark{d}    &  5.5$\times$3.8\tablenotemark{d}      & -15   & --\tablenotemark{e} & MMS 6-NE \\
OMC3-SMM 8	& 5 35 23.42 & -5 01 30.35  & 6152$\pm$44      & 4400$\pm$20    & 3.9$\pm$0.03$\times$1.4$\pm$0.1   	& 73$\pm$1      & 0.86     & MMS 6, CSO 10\\
OMC3-SMM 9	& 5 35 23.49 & -5 01 39.21  & 874$\pm$60       & 3900$\pm$19    & 5.2$\pm$0.3$\times$4.8$\pm$0.3     	& 150$\pm$28    & --\tablenotemark{e} &  \\
OMC3-SMM 10	& 5 35 28.20 & -5 03 40.65  & 220$\pm$53       & 118$\pm$20     & 5.1$\pm$1.1$\times$3.2$\pm$1.2       	& 140$\pm$23    & 0.28     & MMS 7-NE    \\
OMC3-SMM 11	& 5 35 26.56 & -5 03 55.04  & 1162$\pm$41      & 921$\pm$20     & 3.2$\pm$0.2$\times$1.0$\pm$0.4     	& 52$\pm$4      & 0.56     & MMS 7, CSO 12\\
OMC3-SMM 12	& 5 35 25.44 & -5 03 53.22  & 290$\pm$64       & 117$\pm$19     & 7.3$\pm$1.2$\times$3.8$\pm$1.2      	& 80$\pm$12     & 0.32     &              \\
\enddata
\tablenotetext{a}{The source positions were determined from the non-self calibration images with emission peak positions. 
The positional difference between the non self-calibrated images and self-calibrated images is $\approx$0.1$''$.}
\tablenotetext{b}{The flux concentrated ratio is calculated from the total flux measured from the SMA image divided by the JCMT flux within 14$''$ beam.
SMA image are convolved with the JCMT beam size of 14$''$. The uncertainties of the derived ratios are $\pm$0.04 (in absolute value).}
\tablenotetext{c}{REFERENCE.---MMS sources by Chini et al. (1997); CSO sources by Lis et al. (1998); MMS 7-NE by Takahashi et al. (2006); MMS 6-NE by Takahashi et al. (2009)}
\tablenotetext{d}{Flux information and deconvolution size of SMM 6-NE are referred from Takahashi et al. (2009)}
\tablenotetext{e}{Single-dish data taken with the JCMT were not able to fit with the Gaussian components due to the separation between 
sources compared with the large spatial resolution.}
\end{deluxetable}

\begin{deluxetable}{lccccc}
\tabletypesize{\scriptsize}
\tablecaption{Physical Parameters\label{tbl4}}
\tablewidth{0pt}
\tablehead{
\colhead{ID} & \colhead{$M_{H2}$\tablenotemark{a}}  & \colhead{$N_{H_2}$\tablenotemark{a}} & \colhead{$n_{H_2}$\tablenotemark{a}} & \colhead{Source Size\tablenotemark{b}} &  
\colhead{NN distance\tablenotemark{c}} \\
\colhead{} & \colhead{($M_{\odot}$)} & \colhead{(10$^{23}$ cm$^{-2}$)} & \colhead{(10$^{7}$ cm$^{-3}$)} & \colhead{(AU)} & \colhead{(arcsec)} \\
}
\startdata
SMM 1	& 0.5$\pm$0.1  & 0.9$\pm$0.4  	& 0.3$\pm$0.2   	& 3520$\times$950  &  33.3 \\
SMM 2	& 1.6$\pm$0.3  & 17.0$\pm$4.4   & 15.0$\pm$4.5      	& 1160$\times$910  &  15.8 \\
SMM 3	& 1.5$\pm$0.3  & 13.0$\pm$3.1   & 9.6$\pm$2.8     	& 1620$\times$410  &  15.8 \\
SMM 4	& 1.5$\pm$0.3  & 1.4$\pm$0.7   	& 0.4$\pm$0.2   	& 3930$\times$3150 &  21.2 \\
SMM 5	& 3.9$\pm$0.8  & 1.4$\pm$0.4   	& 0.2$\pm$0.1   	& 7870$\times$2150 &  28.7 \\
SMM 6	& 1.9$\pm$0.4  & 11.0$\pm$2.5   & 7.1$\pm$1.8     	& 1620$\times$1160 &  21.3 \\
SMM 7	& 2.4$\pm$0.5  & 7.3$\pm$1.5    & 3.3$\pm$0.7    	& 2280$\times$1570 &  ---  \\
SMM 8	& 5.7$\pm$1.1\tablenotemark{d}  & 44.0$\pm$8.9\tablenotemark{d} & 33$\pm$6.6\tablenotemark{d} & 1620$\times$580 & 8.9 \\
SMM 9	& 1.3$\pm$0.3  & 3.6$\pm$0.9   	& 1.5$\pm$0.4    	& 2150$\times$1990 & 8.9 \\
SMM 10	& 0.3$\pm$0.1  & 1.2$\pm$0.8   	& 0.6$\pm$0.5    	& 2110$\times$1330 & 28.4 \\
SMM 11	& 1.8$\pm$0.4  & 21.0$\pm$5.3   & 19.0$\pm$5.8      	& 1330$\times$410  & 16.8 \\
SMM 12	& 0.4$\pm$0.1  & 0.9$\pm$0.4  	& 0.3$\pm$0.2   	& 3020$\times$1570 & 16.8 \\
\enddata
\tablenotetext{a}{Measurement errors of deconvolved size and total flux described in Table 3 are taken into account for error estimations.}
\tablenotetext{b}{The error for source sizes and position angles are summarized in Table 3 column 5--7, respectively.}
\tablenotetext{c}{The nearest neighbor distance for each source (Section 4.2.1). SMM 7 was excluded from the nearest neighbor calculation since the separation between 
SMM 7 and SMM 8 is much smaller than angular resolution of this paper (4$''$.5). 
We consider that the separation between SMM 7 and SMM 8 are related to formation of a binary system.}
\tablenotetext{d}{$\beta$=0.93 derived by Takahashi et al. (2009) was adopted to estimate the parameters.}
\end{deluxetable}

\begin{deluxetable}{lccccccc}
\tabletypesize{\scriptsize}
\tablecaption{Comparison of Multi-wavelength Data Sets toward Detected Continuum Sources}
\tablewidth{0pt}
\tablehead{
\colhead{ID} & \colhead{CO outflow}  & \colhead{H$_2$ jet} & \colhead{NIR nebula} & \colhead{3.6 cm emission} & \colhead{X-ray} &\colhead{Infrared Source Detection} &\colhead{Evolutionary Status
\tablenotemark{a}} \\
}
\startdata
SMM 1	&  ...     		& ...         		&  ...          & ...        		& ...    		& 2.2, 3.8, 4.5, 5.8, 8, 24$^{\it 9, 10}$ 		& Protostellar core \\
SMM 2	&  ...    		& ...          		& ...           & ...         		& ... 			& ... 						& Prestellar core \\
SMM 3	&  Yes$^{\it 1,2}$  	& Flow B$^{\it 5}$ 	& Yes$^{\it 2}$   & VLA 1$^{\it 6,7}$	& Yes$^{\it 8}$ 	        & 2.2, 3.8, 4.5, 5.8, 8.0, 24$^{\it 9, 10}$ 		& Protostellar core \\
SMM 4	&  ...    		& ...          		& ...    	& ...        		& Yes$^{\it 8}$ 	        & 4.5, 5.8, 8, 24$^{\it 9, 10}$  			& Protostellar core  \\
SMM 5	&  ...    		& ...          		& ...    	& ...        		& ... 			& ... 						& Prestellar core \\
SMM 6	&  Yes$^{\it 1,2}$  	& Flow C$^{\it 5}$  	& Yes$^{\it 2}$   & ...        		& ... 			& 3.6, 4.5, 5.8, 8, 24$^{\it 9, 10}$ 		& Protostellar core \\
SMM 7	&  ...    		&  ...         		& Yes$^{\it 2}$   & VLA 3$^{\it 6}$	& ... 			& 2.2, 3.8, 4.5, 5.8, 8.0, 24$^{\it 9, 10}$ 		& Protostellar core \\
SMM 8	&  Yes$^{\it 3}$    	&  ...         		& ...    	& ... 			& ... 			& 24$^{9, 10}$?$^{b}$ 					& Protostellar core  \\
SMM 9	& ...  			& ...           	& Yes$^{\it 2}$   & ... 			& ... 			& 2.2, 3.6, 4.5, 5.8, 8, 24$^{\it 9, 10}$		& Protostellar core \\
SMM 10	&  Yes$^{\it 4}$   	& ...            	&  Yes$^{\it 2}$  & ...          	& ...   		& 3.6, 4.5, 5.8, 8.0, 24$^{\it 9, 10}$ 		& Protostellar core  \\
SMM 11	& Yes$^{\it 1,2,4}$  	& Flow F$^{\it 5}$   	& Yes$^{\it 2}$   & VLA 4$^{\it 6}$        & ... 			& 2.2, 3.6, 4.5, 5.8, 8, 24$^{\it 9, 10}$ 		& Protostellar core \\
SMM 12	& ...   		& ...            	&  ...  	& ...          		& ...     		& ... 						& Prestellar core \\
\enddata
\tablenotetext{~}{REFERENCE.--- 1: Willams et al. (2003), 2: Takahashi et al. (2008), 3:Takahashi \& Ho (2012), 4: Takahashi et al. (2006), 
5: Yu et al. (1997), 6: Reipurth et al. (1999), 7: Tsujimoto et al. (2004), 8: Tsuboi et al. (2000), 9: Peterson \& Megeath et al. (2008), 10: Megeath et al. (2012)}
\tablenotetext{a}{Evolutionary status of each source was referred from Megeath et al. (2012).}
\tablenotetext{b}{Specially unresolved source at 24 $\mu$m.}
\end{deluxetable}

\clearpage

\begin{deluxetable}{lccccccl}
\rotate
\tabletypesize{\scriptsize}
\tablecaption{Hierarchical Structure of the OMC filaments}
\tablewidth{0pt}
\tablehead{
\colhead{} & \colhead{$n_{\rm{H_2}}$} & \colhead{${T_{\rm{gas}}}$}  & \colhead{Measured Separation} & \colhead{Corresponding Size Scale} \\
\colhead{} & \colhead{(cm$^{-3}$)}  & \colhead{(K)}  &  \colhead{(arcminute)} &\colhead{} \\
}
\startdata
ISMs						&   50--100$^{\it 1}$   	 		& 50--100$^{\it 1}$    & ---                                    	& ISMs observed in HI 21 cm observations  \\
GMCs					&   200-3000$^{\it 2}$  			& ---          			& 282$^{\it 3}$   	   		& Ori A and Ori B (Separation between Orion BN/KL and NGC 2024)\\
Large-scale Clumps			&   9.0e03--3.0e04$^{\it 4}$   	& ---	       			& 9--10$^{\it 5, 6, 7}$         &Significant star forming regions such as OMC1,2,3, etc. \\
Small-scale Clumps                    &   8.8e04--1.1e06$^{\it 5, 8}$ 	& 15--28$^{\it 8}$      & 2.5$^{\it 5}$    		& identified with single-dish NH$_3$ and continuum observations  \\
Dense Cores				&   3.2e06--3.3e08$^{\it 9}$	     	& ---                   		& 0.15--0.55$^{\it 9}$       	& Individual star forming sites detected with the SMA observations \\
Binaries					&   ---	     					& --                               & 0.0025--0.025$^{\it 10}$   & Binary separation range measured in ONC (H$\alpha$ observations).  \\
\enddata
\tablenotetext{~}{REFERENCE.--- 1: Ferriere (2001), 2: Sakamoto et al. (1994), 3: Wilson et al. (2005), 4: Dutrey et al. (1993), 5: Johnstone \& Bally (1999), 6: Dutrey et al. (1991), 7: Hanawa et al. (1993), 8: Cesaroni \& Wilson (1994), 9: This work, 10: Reipurth et al. (2007)}     
\end{deluxetable}


\clearpage
			
\begin{thebibliography}{}
\bibitem[Allen et al.(2007)]{2007prpl.conf..361A} Allen, L., et al.\ 2007, Protostars and Planets V, 361 
\bibitem[Andre et al.(1993)]{1993ApJ...406..122A} Andre, P., Ward-Thompson, D., \& Barsony, M.\ 1993, \apj, 406, 122 
\bibitem[Anglada et al. (1998)]{ang98} Anglada, G., Villuendas, E., Estalella, R., et al. 1998, AJ, 116, 2953
\bibitem[Aso et al.  (2000)]{aso00} Aso, Y., Tatematsu, K., Sekimoto, Y., et al. 2000, \apjs, 131, 465
\bibitem[Bachiller \& Tafalla (1999)]{bac99} Bachiller, R., \& Tafalla, M. 1999, in the Origin of Stars and Planetary Systems, ed, C.J. Lada and N. D. Kylafis (Dordrect: Kluwer), 227
\bibitem[Bally et al. (1987)]{bal87} Bally, J., Langer, W. D., Stark, A. A., \& Wilson, R. W. 1987, \apj, 312, L45
\bibitem[Bate(1998)]{1998ApJ...508L..95B} Bate, M.~R.\ 1998, \apjl, 508, L95 
\bibitem[Beichman et al.(1986)]{1986ApJ...307..337B} Beichman, C.~A., Myers, P.~C., Emerson, J.~P., et al.\ 1986, \apj, 307, 337 
\bibitem[Binney (1987)]{bin87} Binney, J. M., \& Tremaine, S. 1987, Galactic dyamics
\bibitem[Bonnell\& Bate(1994)]{1994MNRAS.269L..45B} Bonnell, I.~A., \& Bate, M.~R.\ 1994, \mnras, 269, L45 
\bibitem[Boss(1989)]{1989ApJ...346..336B} Boss, A.~P.\ 1989, \apj, 346, 336 
\bibitem[Buckle et al.(2012)]{2012MNRAS.422..521B} Buckle, J.~V., Davis, 
C.~J., Francesco, J.~D., et al.\ 2012, \mnras, 422, 521 
\bibitem[Burkert\& Hartmann(2004)]{2004ApJ...616..288B} Burkert, A., \& Hartmann, L.\ 2004, \apj, 616, 288 
\bibitem[Castets et al. (1990)]{cas90} Castets, A., Duvert, G., Dutrey, A., et al. 1990, A\&A, 234, 469
\bibitem[Cesaroni et al. (1994)]{ces94} Cesaroni, R., \& Wilson, T. L. 1994, A\&A, 281, 209
\bibitem[Chini et al. (1997)]{chi97} Chini, R,.	Ward-Thompson, D,.	Bally, J., et al. 1997, \apj,	474, L135
\bibitem[Dutrey (1991)]{dut91} Dutrey, A., Langer, W. D., Bally, J., et al. 1991, A\&A, 247, L9
\bibitem[Dutrey (1993)]{dut93} Dutrey, A., Duvert, G., Castets, A., et al. 1993, A\&A, 270, 468
\bibitem[Evans (1999)]{Eva99} Evans, Neal J., II 1999, ARA\&A, 37, 311
\bibitem[Ferri{\`e}re(2001)]{2001RvMP...73.1031F} Ferri{\`e}re, K.~M.\ 2001, Reviews of Modern Physics, 73, 1031 
\bibitem[Genzel et al. (1989)]{gen89} Genzel, R., \& Stutzki, J. 1989, ARA\&A, 27, 41
\bibitem[Goodwin et al.(2007)]{2007prpl.conf..133G} Goodwin, S.~P., Kroupa, P., Goodman, A., \& Burkert, A.\ 2007, Protostars and Planets V, 133 
\bibitem[Hanawa et al. (1993)]{han93} Hanawa, T., Nakamura, F., Matsumoto T., Nakano, T., et al. 1993, \apj, 404, L83 
\bibitem[Hanawa et al. (1994)]{han94} Hanawa, T., Yamamoto, S., \& Hirahara, Y. 1994, \apj, 420, 318
\bibitem[Harju et al. (1991)]{har91} Harju, J., Walmsley, C. M., \& Wouterloot, J. G. A., 1991, A\&A, 245, 643
\bibitem[Hirota et al. (2007)]{hir07} Hirota, T., Bushimata, T., Choi, Y. K., et al. 2007, PASJ, 59, 897 
\bibitem[Ho et al. (2004)]{ho04} Ho, P. T. P., Moran, J. M., \& Lo, K. Y. 2004, \apj, 616L, 1
\bibitem[Ikeda et al. (2007)]{ike07} Ikeda, N., Sunada, K., \& Kitamura, Y. 2007, \apj, 665, 1194
\bibitem[Jackson et al.(2010)]{2010ApJ...719L.185J} Jackson, J.~M., Finn, S.~C., Chambers, E. T., et al. 2010, \apjl, 719, L185 
\bibitem[Jeans(1902)]{1902RSPTA.199....1J} Jeans, J.~H.\ 1902, Royal Society of London Philosophical Transactions Series A, 199, 1 
\bibitem[Johnstone \& Balley (1999)]{joh99} Johnstone, D., \& Bally, J. 1999, \apj, 510, L49
\bibitem[J$\o$rgensen et al. (2006)]{jor06} J$\o$rgensen, J. K., Johnstone, D., van Dishoeck, E. F., \& Doty, S. D. 2006, A\&A, 449, 609 	 
\bibitem[Keene et al.  (1982)]{kee82} Keene, J., Hildebrand, R. H., \& Whicomb, S. E. 1982, \apj, 252, L11
\bibitem[Kim et al. (2008)]{kim08} Kim, M., Hirota, T., Honma, M., et al. 2008, PASJ, 60, 991 
\bibitem[Lada (2003)]{lad03} Lada, C. J., \& Lada, E. A. 2003, ARA\&A, vol. 41, pp.57--115
\bibitem[Larson(1981)]{1981MNRAS.194..809L} Larson, R.~B.\ 1981, \mnras, 194, 809 
\bibitem[Larson(2002)]{2002MNRAS.332..155L} Larson, R.~B.\ 2002, \mnras, 332, 155 
\bibitem[Lis et al. (1998)]{lis98} Lis, D.C., Serabyn, E., Keene, J., et al. 1998, \apj, 509, 299
\bibitem[Machida et al.(2008)]{2008ApJ...677..327M} Machida, M.~N., Tomisaka, K., Matsumoto, T., \& Inutsuka, S.-i.\ 2008, \apj, 677, 327 
\bibitem[Maddalena et al. (1986)]{mad86} Maddalena, R. J., Morris, M., Moscowitz, J., \& Thaddeus, P. 1986, \apj, 303, 375
\bibitem[Mann et al. (2009a)]{man09a} Mann, R. K., \& Williams, J. P. 2009a, 694, L36 (Mann \& Williams 2009a)
\bibitem[Mann et al. (2009b)]{man09b} Mann, R. K., \& williams, J. P., 2009b, 699, L55 (Mann \& Williams 2009b)
\bibitem[Matsumoto et al. (1994)]{mat94} Matsumoto, T., Nakamura, F., \& Hanawa, T. 1994, \apj, 46, 243
\bibitem[Matthews et al. (2001)]{mat01} Matthews, B. C.\& Willson, C. 2001, \apj, 562, 400
\bibitem[Matthews et al.(2005)]{2005ApJ...626..959M} Matthews, B.~C., Lai, S.-P., Crutcher, R.~M., \& Wilson, C.~D.\ 2005, \apj, 626, 959 
\bibitem[Megeath et al.(2012)]{2012arXiv1209.3826M} Megeath, S.~T., Gutermuth, R., Muzerolle, J., et al.\ 2012, arXiv:1209.3826 
\bibitem[Menten et al. (2007)]{men07} Menten, K. M., Reid, M. J., Forbrich, J., \& Brunthaler, A. 2007, A\&A, 474, 515
\bibitem[Myers et al. (2000)]{mye00} Myers, P. C., Evans, N. J., II \& Ohashi, N., 2000, Protostars and Planets IV (Tucson: Univ. Arizona Press) 
\bibitem[Myers (2009)]{mye09} Myers, P. C. 2009, \apj, 700, 1609
\bibitem[Nakamura et al. (1993)]{nak93} Nakamura, F., Hanawa, T., \& Nakano, T. 1993, PASJ, 45, 551
\bibitem[Nakano\& Tademaru(1972)]{1972ApJ...173...87N} Nakano, T., \& Tademaru, E.\ 1972, \apj, 173, 87 
\bibitem[Nielbock et al. (2003)]{nie03} Nielbock, M., Chini, R., \& Muller, S. A. H. 2003, A\&A, 408, 245
\bibitem[Nutteret al. (2007)]{nut07} Nutter, D., \& Ward-Thompson, D. 2007, MNRAS, 374, 1413
\bibitem[O'Dell et al.(2008)]{2008hsf1.book..544O} O'Dell, C.~R., Muench, A., Smith, N., \& Zapata, L.\ 2008, Handbook of Star Forming Regions, Volume I, 544 
\bibitem[Peterson \& Megeath (2008)]{pet08b} Peterson, D. E., \& Megeath, T. 2008, Handbook of Star Forming Regions, Volume I: The Northern Sky ASP Monograph 
Publications, Vol. 4 Edited by Bo Reipurth, p.59
\bibitem[Pfalzner(2004)]{2004ApJ...602..356P} Pfalzner, S.\ 2004, \apj, 602, 356 
\bibitem[Pfalzner et al.(2005)]{2005A&A...437..967P} Pfalzner, S., Vogel, P., Scharw{\"a}chter, J., \& Olczak, C.\ 2005, \aap, 437, 967 
\bibitem[Pon et al.(2011)]{2011ApJ...740...88P} Pon, A., Johnstone, D., \& Heitsch, F.\ 2011, \apj, 740, 88 
\bibitem[Pon et al.(2012)]{2012ApJ...756..145P} Pon, A., Toal{\'a}, J.~A., Johnstone, D., et al.\ 2012, \apj, 756, 145 
\bibitem[Porras et al. (2003)]{por03} Porras, A., Christopher, M., Allen, L., et al. 2003 \aj, 126, 1916
\bibitem[Pudritz \& Fiege(2000)]{2000ESASP.445..171P} Pudritz, R.~E., \& Fiege, J.~D.\ 2000, Star Formation from the Small to the Large Scale, 445, 171 
\bibitem[Reipurth et al. (1999)]{rei99}Reipurth, B., Rodriguea, L. F., \& Chini R. 1999, \apj,  118, 983
\bibitem[Reipurth (2007)]{Rei07} Reipurth, B., Guimaraes, M. M., Connelley, M. S., \& Bally, J. 2007, AJ, 134, 2272 
\bibitem[Reynolds. (1986)]{rey86} Reynolds, L. F. 1986, \apj, 304, 713
\bibitem[Saigo\& Tomisaka(2006)]{2006ApJ...645..381S} Saigo, K., \& Tomisaka, K.\ 2006, \apj, 645, 381 
\bibitem[Sakamoto et al.(1994)]{1994ApJ...425..641S} Sakamoto, S., Hayashi, M., Hasegawa, T., Handa, T., \& Oka, T.\ 1994, \apj, 425, 641 
\bibitem[Sandstrom et al. (2007)]{san07} Sandstrom, K. M., Peek, J. E. G., Bower, G. C., Bolatto, A. D., \& Plambeck R. L. 2007, \apj, 667, 1161
\bibitem[Scoville et al. (1993)]{sco93} Scoville, N. Z., Carlstrom, J. E., Chandler, C. J., et al. 1993, \apj, 105, 1482
\bibitem[Shimajiri et al. (2008)]{shi08} Shimajiri, Y., Takahashi, S., Takakuwa, S., Saito M., \& Kawabe, R. 2008, \apj,  683, 255
\bibitem[Shimajiri et al. (2009)]{shi09} Shimajiri, Y., Takahashi, S., Takakuwa, S., Saito M., \& Kawabe, R. 2009, PASJ, 61, 1055
\bibitem[Shu(1977)]{1977ApJ...214..488S} Shu, F.~H.\ 1977, \apj, 214, 488 
\bibitem[Solomon et al.(1987)]{1987ApJ...319..730S} Solomon, P.~M., Rivolo, A.~R., Barrett, J., \& Yahil, A.\ 1987, \apj, 319, 730 
\bibitem[Stodolkiewica (1963)]{sto63} Stodolkiewicz, J. S. 1963, Acta Astron., 13(1), 30
\bibitem[Stanke et al. (2002)]{sta02} Stanke, T., McCaughrean, M. J., Zinnecker H. 2002,  A\&A, 392, 239
\bibitem[Takahashi et al. (2006)]{tak06} Takahashi, S., Saito, M., Takakuwa S., \& Kawabe, R. 2006, \apj, 651, 933
\bibitem[Takahashi et al. (2008a)]{tak08a} Takahashi, S., Saito, M., Takakuwa, S., \& Kawabe, R. 2008, Ap\&SS, 313, 165 (Takahashi et al. 2008a)
\bibitem[Takahashi et al. (2008b)]{tak08b} Takahashi, S., Saito, M., Ohashi, N., et al. 2008, \apj, 688, 344 (Takahashi et al. 2008b)
\bibitem[Takahashi et al. (2009)]{tak09} Takahashi, S., Ho, T. P. T., Tang, Y.-W., Kawabe, R., \& Saito, M., 2009, \apj, 704, 1459
\bibitem[Takahashi \& Ho(2012)]{2012ApJ...745L..10T} Takahashi, S., \& Ho, P.~T.~P.\ 2012, \apjl, 745, L10 
\bibitem[Takahashi et al.(2012)]{2012ApJ...752...10T} Takahashi, S., Saigo, K., Ho, P.~T.~P., \& Tomida, K.\ 2012, \apj, 752, 10 
\bibitem[Tatematsu (1993)]{tat93} Tatematsu, K., Umemoto, T., Kameya, et al. 1993, \apj, 404, 643	
\bibitem[Tatematsu et al.(2008)]{2008PASJ...60..407T} Tatematsu, K., Kandori, R., Umemoto, T., \& Sekimoto, Y.\ 2008, \pasj, 60, 407 
\bibitem[Teixeira (2012)]{tei10}	Teixeira, P. S., et al. 2012 in prep.
\bibitem[Tsuboi et al.(2001)]{2001ApJ...554..734T} Tsuboi, Y., Koyama, K., Hamaguchi, K., et al.\ 2001, \apj, 554, 734 
\bibitem[Tsujimoto et al.(2004)]{2004PASJ...56..341T} Tsujimoto, M., Koyama, K., Kobayashi, N., et al.\ 2004, \pasj, 56, 341 
\bibitem[Tsujimoto (2005)]{tuj05} Tsujimoto, M., Feigelson, E. D., Grosso, N., et al. 2005, \apj, 160, 503 	
\bibitem[Vicente. (2005)]{vic05} Vicente, S. M., \&Alves, J. 2005, A\&A, 441, 195
\bibitem[Williams et al. (2003)]{wil03} Williams, J. P., Plambeck, R. L., \& Heyer, M. H. 2003, \apj,  591, 1025
\bibitem[Wilner et al. (1994)]{wil94} Wilner, D. J., \& Welch, W. J., 1994, \apj, 427, 898
\bibitem[Wilson et al.(2005)]{2005A&A...430..523W} Wilson, B.~A., Dame, T.~M., Masheder, M.~R.~W., \& Thaddeus, P.\ 2005, \aap, 430, 523 
\bibitem[Wiseman \& Ho (1998)]{wis98} Wiseman, J. J., \& Ho P. T. P. 1998, \apj,  502, 676
\bibitem[Wright \& Sault (1993)]{wis93} Wright, M. C. H., \& Sault, R. J. 1993, \apj, 402, 546
\bibitem[Yu et al.(1997)]{1997ApJ...485L..45Y} Yu, K.~C., Bally, J., \& Devine, D.\ 1997, \apjl, 485, L45 
\bibitem[Zapata et al. (2004)]{zap04} Zapata, L. A., Rodriguez, L. F., Kurtz, S. E., O'Dell C. R., \& Ho, P. T. P.,  2004  \apj,  610, 121
\bibitem[Zapata et al. (2005)]{zap05} Zapata, L. A., Rodriguez, L. F., Ho, P. T. P., et al. 2005  \apj,  630, 85
\bibitem[Zapata et al. (2006)]{zap06} Zapata, L. A., Ho, P. T. P., Rodriguez, L. F., et al. 2006 \apj,  653, 398
\bibitem[Zapata et al. (2007)]{zap0} Zapata, L. A., Ho, P. T. P., Rodriguez, L. F., Schilke, P., \& Kurtz, S. 2007 \apj,  L471, 59


\end{thebibliography}
\end{document}